\documentclass[fleqn,usenatbib]{mnras}

\usepackage{newtxtext,newtxmath}

\usepackage[T1]{fontenc}
\usepackage{ae,aecompl}

\usepackage{graphicx}	
\usepackage{amsmath}	
\usepackage{amssymb}	
\usepackage{gensymb}
\usepackage{float}
\usepackage{rotating}
\usepackage{multirow}
\usepackage{ulem}
\usepackage{longtable}

\newcommand{\HI}{H\,{\sc i}}
\newcommand{\Amod}{$A_\mathrm{mod}$}
\newcommand{\Aflux}{$A_\mathrm{flux}$}
\newcommand{\cm}{$\mathrm{cm}^{-2}$}

\title[Effect of observational constraints on \HI\ asymmetries.]{Using \textsc{eagle} simulations to study the effect of observational constraints on the determination of \HI\ asymmetries in galaxies}

\author[P. V. Bilimogga et al.]{
Pooja V. Bilimogga,$^{1}$\thanks{E-mail: pooja@astro.rug.nl}
Kyle A. Oman$^{1,2,3}$
Marc A.W. Verheijen,$^{1}$
Thijs van der Hulst,$^{1}$
\\
$^{1}$Kapteyn Astronomical Institute, University of Groningen, Postbus 800, 9700 AV Groningen, The Netherlands\\
$^{2}$Department of Physics, Durham University, South Road, Durham DH1 3LE, United Kingdom\\
$^{3}$ Institute for Computational Cosmology, Durham University, South Road, Durham DH1 3LE, United Kingdom
}

\date{Accepted XXX. Received YYY; in original form ZZZ}

\pubyear{2021}

\begin{document}
\label{firstpage}
\pagerange{\pageref{firstpage}--\pageref{lastpage}}
\maketitle

\begin{abstract}
We investigate the effect of observational constraints such as signal-to-noise, resolution and column density level on the \HI\ morphological asymmetry (\Amod) and the effect of noise on the \HI\ global profile (\Aflux) asymmetry indices. Using mock galaxies from the \textsc{eagle} simulations we find an optimal combination of the observational constraints that are required for a robust measurement of the \Amod\ value of a galaxy: a column density threshold of $5\times10^{19}$\cm\ or lower at a minimal signal-to-noise of 3 and a galaxy resolved with at least 11 beams. We also use mock galaxies to investigate the effect of noise on the \Aflux\ values and conclude that a global profile with signal-to-noise ratio greater than 5.5 is required to achieve a robust measurement of asymmetry. We investigate the relation between \Amod\ and \Aflux\ indices and find them to be uncorrelated which implies that \Aflux\ values cannot be used to predict morphological asymmetries in galaxies.
\end{abstract}




\section{Introduction}
\par Characterisation of galaxies has been a fundamental part of astronomy for over a century. The morphology of many nearby galaxies have been characterised based on visual structures and features and subsequently classified into different categories. One of the earliest schemes of classification was introduced by \cite{Hubble1926} who classified galaxies into ellipticals and spirals with further classification based on the presence of bars. This was later refined into the `Tuning Fork' scheme by \cite{Sandage1961}. Schemes of classification that are in use today were developed by \cite{deVaucouleurs1959} \cite{vdBergh1960,vdBergh1976}, and \cite{Elmegreen1987} among others. These schemes are an extension of the `tuning fork' scheme and classify galaxies on the basis of prominent visual features such as bars, rings, the winding of spiral arms and clumpiness of light in these arms. Physical properties of galaxies such as colour, stellar mass, and star formation rate correlate with their morphological type \citep{Holmberg1958, Roberts1994,Conselice2006}. Moreover, the morphological type of galaxies also correlates with the local environmental density  \citep{Dressler1980, Dressler1984}. These correlations offer important clues about the underlying physics of galaxy formation.
\par As opposed to the visual and qualitative classification schemes, quantitative techniques have also been utilised to classify galaxies. Detailed investigations into the light profiles of elliptical galaxies was introduced by \cite{deVaucouleurs1948} and later generalised for other morphological types by \cite{Sersic1963}. Light profiles of galaxies were decomposed into bulge and disc profiles by \cite{Kormendy1977, Kormendy1977b} in addition to other complex features such as bars, lenses, and inner and outer rings \citep{Kormendy1979}. Such detailed quantitative studies were performed by \cite{DeJong1996, DeJong1996b} for spiral galaxies and by \cite{Kormendy2009} for elliptical galaxies, taking advantage of charge-couple device (CCD) imaging.
\par Light structures in galaxies are also characterised using non-parametric methods. Using images from the Hubble space telescope, \cite{Schade1995} measured the degree of peculiarity in galaxies and \cite{Abraham1996} characterised asymmetry and concentration of distant galaxies to objectively classify their morphologies. \cite{Conselice1997} characterised the asymmetry in nearby galaxies and found a correlation between the asymmetry values of the galaxies and their B-V color such that bluer galaxies are more asymmetric. Further investigation by \cite{Conselice2000} revealed that the asymmetry-color correlation is useful for classifying distant galaxies into different morphological types as well as identifying interacting systems. Together, concentration, asymmetry, and smoothness are collectively referred to as the CAS parameters and are the most common non-parametric methods to characterise galaxy structure \citep{Conselice2003}. Other parameters such as Gini and M20 have been used, in combination with the CAS parameters, to identify ultra-luminous infrared galaxies (ULIRGs) and on-going mergers \citep{Abraham2003, Lotz2004}.
\par Various dynamical processes that act on galaxies not only distort the stellar distribution but also the distribution of gas. The kinematically cold, collisional, and often extended neutral atomic hydrogen (\HI) disc in the outer parts of galaxies are fragile in nature and easily disturbed by external processes even before the stellar disc is affected. This makes \HI\ an excellent tracer of environmental processes such as tidal interactions with neighbouring galaxies, mergers, and interactions with the surrounding gaseous medium. For instance, tidal interactions with a neighbouring galaxy can be inferred from long, trailing streams of gas (for example NGC 4111 and NGC 4026 mentioned in \cite{Verheijen2001Lenticulars}). When galaxies merge, complex \HI\ features such as bridges and plumes of neutral gas may be seen around the merging system (references mentioned in \cite{Sancisi1999}). Gas-rich galaxies entering a dense cluster environment interact with the surrounding hot intra-cluster medium resulting in \HI\ distributions offset from their stellar disc due to ram pressure \citep{Chung2009}. Clusters are also known to host galaxies with truncated \HI\ discs, possibly due to harassment, starvation or thermal evaporation \citep{Chung2009, Moore1998, Warmels1988}. Another peculiar feature that is commonly observed in galaxies is a lopsidedness in the \HI\ distribution, the origin of which cannot always be explained as the result of an ongoing tidal interaction but may be the result of minor mergers or past tidal interactions \citep{Sancisi2008}. Peculiar features observed in \HI\ are documented and classified in the \HI\ rogues gallery \citep{HIRogues}, illustrating how the environmental impact on galaxies produces asymmetric and distorted \HI\ distributions.
\par \cite{Baldwin1980} first studied the lopsided distribution of \HI\ gas in $\sim20$ nearby galaxies and proposed a pattern of elliptical orbits in which the gas moves to explain the existence of these features. Thereafter, several techniques have been used to characterise asymmetry in the spatial distribution of the \HI\ gas. A Fourier decomposition method was applied to the column density maps of large and nearby galaxies in the Eridanus group by \cite{Angiras2006} and for the WHISP sample \citep{Whisp2001} by \cite{Eymeren2011}. The CAS, Gini, and M20 indices were applied to \HI\ maps of the WHISP sample by \cite{Holwerda2011} and \cite{Giese2016}. \cite{Lelli2014} applied a modified asymmetry index to study the structure of the \HI\ distribution in irregular star-burst galaxies. \cite{Holwerda2011} find that non-parametric methods only weakly correlate with visual classifications of galaxies in the WHISP sample. On the other hand, \cite{Giese2016} and \cite{Lelli2014} report that the asymmetry parameter is better suited in identifying galaxies with lopsided \HI\ distributions.
\par Environmental effects on the gas disc of a galaxy can also be inferred by characterising the global \HI\ profile asymmetry through a comparison of the flux in the approaching and receding parts of the profile. Using this method, a large number of objects from the single-dish surveys can be analysed to understand the environmental effect with the caveat that a skewed global \HI\ profile can result from either morphological or kinematic lopsidedness. \cite{Richter1994} found that $\sim50$ per cent of a sample of 1700 galaxies showed strong lopsidedness in their global profiles and concluded that lopsidedness in the global profile may be the rule rather than the exception. \cite{Haynes1998} investigated the global profiles of 104 isolated galaxies, of which $\sim50$ per cent show lopsided global profiles. Similarly, \cite{Espada2011} studied high signal-to-noise profiles of 166 extremely isolated galaxies and established an intrinsic asymmetry rate against which galaxies in different environments can be compared. \cite{Scott2018} studied galaxies in Abell 1367 and the Virgo cluster to find that 26 per cent and 16 per cent of galaxies have asymmetries in their global profiles. \cite{Bok2019} investigated the global profiles of $\sim350$ close pairs of galaxies from the ALFALFA sample and found that asymmetries in global profiles are common in close pairs of galaxies. Asymmetries in global profiles have been studied for many large samples of galaxies from which environmental effects have been inferred. Effects of observational parameters such as noise and spectral resolution of the profile on the asymmetries have been quantified by \cite{Watts2020}, who also established a robust procedure for the comparison of profile asymmetries.
\par The application of an asymmetry index to \HI\ column density maps from interferometric observations is a relatively new endeavour and our understanding of the effect of observational parameters on the measured morphological asymmetry values is limited. \cite{Giese2016} have used models of lopsided gas distributions to study how galaxy inclination, noise in the datacube, and the number of beams across a galaxy affects the measured morphological asymmetry value. However, each of these effects was studied separately. Since these aspects are not independent, a detailed investigation of their combined effect on the morphological asymmetry index is required. For this, we use mock \HI\ datacubes of galaxies from the \textsc{eagle} simulations \citep{Schaye2015,Crain2015} smoothed to different spatial resolutions and with a range of noise levels applied. We investigate what combination of constraints on the resolution, column density threshold and noise is required to obtain a robust measurement of asymmetry in the column density maps of galaxies. We also investigate the relation between the global profile and the morphological asymmetry indices. If such a correlation exists, it would make it possible to infer the morphological asymmetry of unresolved galaxies based on their global profile asymmetry. This will be essential for upcoming \HI\ surveys such as WALLABY \citep{Wallaby2020} and MIGHTEE-HI \citep{Mightee2016} as well as deep \HI\ imaging surveys such as LADUMA \citep{Laduma2016} and CHILES \citep{Chiles2013} in which many galaxies beyond the local universe will not be spatially resolved to measure their morphological asymmetry.
\par This paper is organised as follows: Section \ref{sec:Asymmetry} describes the morphological and the global \HI\ profile asymmetry indices, Section \ref{sec:MockSample} describes the sample of mock galaxies from the \textsc{eagle} simulation used in this work, Section \ref{sec:Moprhasymmetry} discusses the effect of observational parameters such as column density threshold, resolution and noise on the morphological asymmetry index. Section \ref{sec:GPasymmetry1} describes the results of global profile asymmetry measurements for the mock sample as well as a verification of the effect of noise on the global profile asymmetry index. In Section \ref{sec:CorrelateAmodAflux} we investigate the relation between global profile and morphological asymmetry indices, and Section \ref{sec:Conclusion} summarises the results of this work.

\section{Asymmetry indices}\label{sec:Asymmetry}
\subsection{Measuring morphological asymmetry}
\par The asymmetry index ($A$) that measures the deviations from symmetry in the distribution of matter in a galaxy is defined as follows:
\begin{equation}\label{Asymm}
    A = \frac{\sum_{i,j} |{I(i,j) - I_{180}(i,j)}|}{2\sum_{i,j} |I(i,j)|}
\end{equation}
where $I$ denotes the original image of the galaxy and $I_{180}$ is the image rotated by $180 \degree$ around a chosen center. The asymmetry value is measured by taking a pixel-by-pixel difference between the original and rotated image, which is then summed and normalised by the total intensity in the image. The asymmetry index can have a value between 0 and 1. Due to the way the asymmetry index is defined, brighter pixels which tend to be near the centre of an image contribute more to the index than the fainter pixels in the outer regions. The influence of environmental processes is likely to disturb the extended outer parts of the \HI\ disc.
\par To ensure an equal contribution of all pixels to the asymmetry index and to measure the contribution from the outer parts of an \HI\ disc, a modification to the index was introduced by \cite{Lelli2014}. The modified asymmetry index (\Amod) is defined as:
\begin{equation}\label{AsyMod}
    A_\mathrm{mod} = \frac{1}{N}\sum_{i,j}^{N} \frac{|{I(i,j) - I_{180}(i,j)}|}{|{I(i,j) + I_{180}(i,j)}|}
\end{equation}
where the residuals are normalized with respect to the `local' intensity as opposed to the total intensity of the pixels. Through this modification, asymmetries in the fainter outer parts can be better measured than with the classic asymmetry index.
\par The measured values of asymmetry are biased by the effect of noise and thus should be interpreted with caution especially when the signal-to-noise ratio is low. To correct for the effects of noise, \cite{Conselice2000} calculate the bias in the asymmetry value from emission-free, i.e. blank, regions of an image. The corrected asymmetry values are then calculated as follows:
\begin{equation}
    A_\mathrm{corrected} = \frac{\sum_{i,j} |{I(i,j) - I_{180}(i,j)}|}{2\sum_{i,j} |I(i,j)|} - \frac{\sum_{i,j} |{B(i,j) - B_{180}(i,j)}|}{2\sum_{i,j} |I(i,j)|}
\end{equation}
where $B$ is the `blank' image derived from applying the mask onto neighboring blank regions of the image and $B_{180}$ is the blank image rotated by $180\degree$. \cite{Giese2016} pointed out that this bias parameter, designed for optical images, does not work well for typical \HI\ distributions as the signal-to-noise is usually much lower and the background noise is on average zero, thus producing an over-correction. \cite{Giese2016} introduced a measure of the signal-to-noise in the difference image and used machine learning methods to evaluate the bias in the measured asymmetry as a function of the signal-to-noise in the \HI\ column density image as well as the signal-to-noise in the difference image. However, noise is not the only source of uncertainty in asymmetry measurements. Factors such as spatial resolution and inclination also affect the measured asymmetry value. The independent effect of these factors is studied  by \cite{Giese2016} using model galaxies. When comparing asymmetry values of galaxies observed with different sensitivities, the choice of column density threshold applied to the maps also becomes important. Resolution and column density are not independent in \HI\ data; lowering the spatial and spectral resolution of a datacube will improve the column density sensitivity. Therefore, an optimum combination of noise level, resolution and column density threshold is needed to measure and compare asymmetry values in an unbiased manner.  In this work we use the \Amod\ index to measure the asymmetry in the \HI\ column density maps of mock galaxies.

\subsection{Measuring asymmetries in the global \HI\ profile} \label{sec:GPAsymmetry}
\par Asymmetries in the global profiles of galaxies are quantified by measuring the ratio of fluxes in the two halves of the emission line. The integrated flux ratio index (\Aflux) is defined as follows:
\begin{equation}\label{Aflux}
    A_\mathrm{flux} = \frac{\int_{V_\mathrm{sys}}^{V_\mathrm{high}} S_\nu dv}{\int_{V_\mathrm{low}}^{V_\mathrm{sys}} S_\nu dv}
\end{equation}
where $V_\mathrm{low}$ and $V_\mathrm{high}$ are the velocities at the 20 per cent level of the peak flux density value of the spectrum and $V_\mathrm{sys}$ is the mid-point between the $V_\mathrm{low}$ and the $V_\mathrm{high}$ velocities. The flux values in the spectrum are linearly interpolated between consecutive channels. If the $V_\mathrm{low}$ and $V_\mathrm{high}$ velocities lie within a channel then a fraction of the flux is integrated in Eq. \ref{Aflux}. If \Aflux$\,<1$ then the reciprocal of the ratio defined in Eq. \ref{Aflux} is considered instead. A symmetric profile will result in \Aflux$=1$ and deviations from symmetry in the spectrum will yield \Aflux$\,>1$.
\par When measuring and interpreting the asymmetry in a global profile one should take into account the effect of noise in the observation as well as the orientation of the galaxy with respect to the observers' line of sight. Using model global profiles, \cite{Watts2020} studied the dependence of \Aflux\ values on the signal-to-noise (S/N) ratios of the global profile. They found that at low S/N ratios, their sample of model galaxies shows higher rates of asymmetry. They also describe a robust method to account for the effect of noise on the \Aflux\ values. \cite{DegHank2020} investigated how \Aflux\ values depend on the viewing angle and inclination using two snapshots of interacting galaxies. They find that despite an underlying morphologically asymmetric gas distribution, the shape of the global profile may be symmetric at certain combinations of viewing angle and inclination. \cite{DegHank2020} infer that an asymmetric global profile indicates an asymmetric gas distribution, however a symmetric global profile need not necessarily indicate symmetry in the underlying gas distribution.

\section{The mock galaxy sample}\label{sec:MockSample}
\par To explore the optimal combination of observational parameters that allow for a robust measurement of asymmetries, these need to be varied over a large range of values. Datacubes from actual \HI\ observations of galaxies are not suited for this exercise as the relevant parameters are observationally interlinked. By using mock \HI\ datacubes from hydrodynamical simulations, parameters such as resolution and noise can be accurately and independently controlled. Low column density levels in spatial \HI\ distribution and different orientations of the disc can be probed. In this work, we use mock \HI\ datacubes of galaxies from the \textsc{eagle} simulations as the simulated galaxies better allow for the complexities of real galaxies to be seamlessly accounted for than model galaxies. Rotational velocities of galaxies, the resolution of the mock \HI\ datacubes, and noise levels used in this work have been selected such that they resemble the interferometric observations of Ursa Major and Perseus-Pisces galaxies to be presented in our forthcoming publication.
\subsection{The \textsc{eagle} simulations}
\par The \textsc{eagle} simulations \citep{Schaye2015,Crain2015} are a suite of cosmological simulations run with a modified \textsc{Gadget-3} N-Body Tree-PM smoothed particle hydrodynamics (SPH) code described in \cite{Springel2005}. The simulations follow the evolution of gas and dark matter over 29 snapshots over the redshift range $z = 0 - 20$, for a range of resolutions and parameters sets for the sub-grid models. The various sub-grid physical models implemented in \textsc{eagle} include the radiative cooling of gas \citep{Wiersma2009a}, reionization \citep{Wiersma2009b, Haardt2001}, star formation \citep{Schaye2008}, stellar mass loss \citep{Wiersma2009b}, energy feedback from star formation \citep{DallaVecchia2012}, and active galactic nuclei (AGN) feedback \citep{Rosas2015}. In the simulations, the model parameters regulating the energy feedback from star formation and AGNs were calibrated to reproduce the observed galaxy stellar mass function (GSMF) at $z \sim 0$. Additionally, a dependence of the stellar feedback energy on the gas density was introduced to reproduce the galaxy mass-size relation at $z \sim 0.1$. A comprehensive description of these calibration procedures can be found in \cite{Crain2015}. The \textsc{eagle} simulations use $\Lambda$CDM cosmological parameters consistent with the Planck 2014 results \citep{PlanckCollab2014}: $\Omega_m = 0.307$, $\Omega_\Lambda = 0.693$, $\Omega_b = 0.04825$, $h = 0.6777$ and $\sigma_8 = 0.8288$.
\par We select our sample of galaxies from the \textsc{RECAL} model run of \textsc{eagle} simulation `RecalL0025N0752' (hereafter Recal25), which is a cosmological volume of (25 Mpc)$^3$ (comoving) with $752^3$ dark matter particles. The model initially has an equal number of dark matter and baryonic particles, with a dark matter particle mass of $1.21\times10^{6}\mathrm{M}_{\odot}$, and an initial baryonic particle mass of $2.26\times10^{5}\mathrm{M}_{\odot}$. In the simulation box, galaxies are defined as gravitationally bound sub-haloes, which are identified using the SUBFIND algorithm \citep{Springel2005, Dolag2009}. Initially, the dark matter particles are grouped into haloes by running the friends-of-friends (FoF) algorithm with a linking length of 0.2 times the mean inter-particle separation. Gas and star particles are assigned to the same FoF halo as their nearest dark matter particle. Within a FoF halo, saddle points in the density distribution are used to define substructures. Any particles that are not bound gravitationally to the identified substructure are removed and resulting substructures are called subhaloes. In each substructure identified by the algorithm, the most massive sub-halo with the lowest value of gravitational potential is defined as a `central galaxy' and remaining sub-halos in the substructure are labelled as `satellite galaxies'. The position of a galaxy is defined by the position of the particle which has the lowest value of its gravitational potential within the subhalo (see \citealt{Schaye2015} for more details).

\subsection{Mock \HI\ datacubes}\label{MockHICubes}
\par We select 189 galaxies from the Recal25 simulation that are central galaxies, i.e. the most massive in their FoF group,  with maximum rotational velocities ($V_\mathrm{max}$) in the range $80\,\mathrm{km}\,\mathrm{s}^{-1}<V_\mathrm{max}<200\,\mathrm{km}\,\mathrm{s}^{-1}$ inferred from the rotation curve. We choose this range to match the $V_\mathrm{max}$ values of galaxies in the Ursa Major volume targeted with the Westerbork Synthesis Radio Telescope (WSRT) as described in Table 3 of \cite{Verheijen2001}. This results in the sample having \HI\ masses in the range $8.08 <$log($M_\mathrm{HI}/M_{\odot}$)$<10.25$, however we do not enforce this in the selection of our sample. To create a mock \HI\ cube, we use MARTINI \footnote{https://github.com/kyleaoman/martini} \citep{MartiniASCL} which is a modular package for the creation of synthetic resolved HI datacubes from SPH simulations of galaxies. The underlying methodology used to create mock \HI\ datacubes is as follows (see also Section 3.3 of \citealt{Oman2019}). Following the prescription of \cite{Rahmati2013}, the neutral hydrogen gas fraction for each gas particle is calculated while accounting for self-shielding from the ionizing background radiation. Thereafter, the molecular component of the gas particle is computed using the empirical pressure-dependent relation by \cite{Blitz2006}. This empirical prescription is used so that the partitions of the atomic and molecular components of the gas particles are computed consistently with other previous works that utilise the \textsc{eagle} simulation (See \cite{Bahe2016}, \cite{Marasco2016}, \cite{Crain2017}). The pressure-dependent fraction of the molecular component of the gas particle is computed as follows:
\begin{equation}
    R_\mathrm{mol} = \left (\frac{P}{P_0}\right)^\alpha
\end{equation}
where $P = n_\mathrm{H}k_\mathrm{B}T$ is the pressure of the gas at temperature T and $P_{0}/k_\mathrm{B} = 4.3\ \times\ 10^{4} \mathrm{cm}^{-3}$ K and $\alpha = 0.92$.
\par Subsequently, the $C^2$ smoothing kernel \citep{Wendland1995} is used to spatially smooth the gas particles. Along each line-of-sight, the 21-cm line emission is modelled with a Gaussian profile centered on the particle velocity with an adaptive broadening depending on the temperature of the particle. The Gaussian thermal velocity  dispersion of a single particle is given by:
\begin{equation}\label{eq:gasdisp}
    \sigma_\mathrm{T} = \sqrt{\frac{k_\mathrm{B}T_\mathrm{g}}{m_\mathrm{p}}}
\end{equation}
where $\sigma_\mathrm{T}$ is the temperature dependent dispersion of the velocity profile, $k_\mathrm{B}$ is the Boltzmann's constant, $T_\mathrm{g}$ is the gas temperature, and $m_\mathrm{p}$ is the proton mass. The typical temperature of the gas particles is about 8000 K, which results in $\sigma_\mathrm{T} = 8.1 \mathrm{km}\,\mathrm{s}^{-1}$ for each particle. The exact value of this thermal velocity dispersion of a single particle, however, does not determine the total line-of-sight velocity dispersion which is dominated by the line-of-sight velocity distribution of multiple particles. This distribution of particles has a typical dispersion of several tens of kilometers per second for the galaxies in our sample.
\par The \HI\ gas is assumed to be optically thin and thus each particle contributes flux proportional to its \HI\ mass. Thereafter, a co-ordinate system is adopted that is centered on the minimum of the galaxy's gravitational potential. The orientation of the galaxy is simply that it has in the intrinsic coordinate system of the simulation, with the line of sight along the first axis (x-axis). We simulate an \HI\ observation by placing the galaxy in the Hubble flow at a distance of 17 Mpc, which is the distance to the Ursa Major volume, thereby setting the angular scale of the mock observation to 82 parsec per arcsecond. The following steps are taken so that the mock \HI\ datacubes are similar to the datacubes of the UMa galaxies obtained with the WSRT (see \citealt{Verheijen2001}). A mock \HI\ datacube is created with $512^2$ pixels, a pixel size of $5\ \mathrm{ arcsec} \times 5\ \mathrm{ arcsec}$, and a channel width of $4\,\mathrm{km}\,\mathrm{s}^{-1}$. Thereafter, we convolve the datacubes to a Gaussian beam of $12\ \mathrm{arcsec} \times 17\ \mathrm{arcsec}$ and also apply a Hanning taper to set the velocity resolution to $8\,\mathrm{km}\,\mathrm{s}^{-1}$. The unit of the pixel values in these mock datacubes is $\mathrm{Jy}\,\mathrm{beam}^{-1}$. At this point no instrumental noise is added to these simulated datacubes.
\par Since the mock datacube for each galaxy is made from the set of particles in the FoF group, there may be contamination from the presence of satellite galaxies that belong to the same FoF group but happen to be directly in the foreground or background. We do not exclude such galaxies from the mock datacube as we endeavour to keep the process as close to real-life observational situations as possible. A 3D mask, which is guided by the rms noise in the mock datacube, is sufficient in excluding distinct foreground/background galaxies from the galaxy of interest in the centre of the datacube.For the noise-free mock datacubes of all 189 galaxies, such chance alignments have been visually inspected and datacubes with satellite galaxies in them are excluded from the analysis.
\par In actual observed datacubes, the column density sensitivity can be improved by smoothing the datacubes to lower angular resolutions. In preparation of future comparisons to actual observed data, the mock datacubes have been further smoothed to angular resolutions of 30, 45, 56, 98 and 120 arcsec. At the distance of 17 Mpc, these angular resolutions correspond to 1.65, 2.47, 3.70, 4.62, 8.07, and 9.89 kpc, respectively. Column density maps are created by adding the pixel values along the velocity axis to obtain a total flux in units of $\mathrm{Jy}\,\mathrm{beam}^{-1}$. This is then converted to column density units of \cm\ by using
\begin{equation}\label{Eq:ColDen}
    N_\mathrm{HI} = 1.83 \times 10^{18} \int T_\mathrm{b} dv
\end{equation}
where $T_\mathrm{b}$ corresponds to the brightness temperature of the emission in Kelvin and $dv$ to the channel width in $\mathrm{km}\,\mathrm{s}^{-1}$. The conversion from $\mathrm{Jy}\,\mathrm{beam}^{-1}$ to brightness temperature $T_\mathrm{b}$ is given by:
\begin{equation}
    T_\mathrm{b} = \frac{605.7}{\Theta_{x}\Theta_{y}} S_{\nu} \bigg( \frac{\nu_{o}}{\nu} \bigg)^2
\end{equation}
where $\Theta_x$ and $\Theta_y$ are major and minor axes of the Gaussian beam in arcsec, $S_{\nu}$ is the flux density in $\mathrm{mJy}\,\mathrm{beam}^{-1}$, and $\nu_{o}$ and $\nu$ are the rest and observed frequency of the \HI\ line emission. For a distance of 17 Mpc, $\frac{\nu_{o}}{\nu} \sim 1$. Global profiles are derived from the mock datacubes by spatially integrating the entire flux in each channel. Asymmetries in these global profiles are measured using Eq. \ref{Aflux} and the morphological asymmetries in the column density maps are measured using Eq. \ref{AsyMod} considering only pixels above the column density thresholds of $1, 2, 5, 10, 15, 20, 45 \times 10^{19}\text{ cm}^{-2}$ for analysis purposes.

\subsection{Mock \HI\ datacubes with added noise}
\par To measure the effect of noise on the asymmetry values, we use the WSRT datacube of UMa galaxy UGC 6805 as the noise cube. UGC 6805 is a dwarf elliptical galaxy in the Ursa Major volume with an undetectable amount of \HI\ gas and there is no strong continuum source present in the datacube making it suitable to be used as a noise cube representative of a real observation. We add this noise cube to the noise-free mock datacubes at $12\ \mathrm{arcsec} \times 17\ \mathrm{arcsec}$ resolution. Next, the noise-added mock \HI\ datacubes are smoothed to lower angular resolutions with circular beams of 30, 45, and 56 arcsec and a velocity resolution of $20\ \mathrm{km}\,\mathrm{s}^{-1}$. The noise at the highest resolution is scaled such that column densities of $1, 2, 5, 10, 15, \text{ and } 20 \times 10^{19}\text{ cm}^{-2}$ have a signal-to-noise ratio of 1, 2, 3, 4, and 5 at the different angular resolutions. This results in 120 mock \HI\ datacubes for every galaxy selected from Recal25, each with a different combination of resolution and noise level. 
\par Each of the 120 noise-added mock datacubes were smoothed to 98 arcsec and $40\,\mathrm{km}\,\mathrm{s}^{-1}$ resolution in order to define a 3D mask to isolate the \HI\ emission.  We applied a clip level of 2.5$\sigma$ to create a 3D mask where pixels below the clip level are set to zero and above to one. This 3D mask is then applied to the corresponding higher resolution datacube and column density maps are created using Eq. \ref{Eq:ColDen}. Signal-to-noise (S/N) ratios corresponding to a column density level are measured according to the method described in \cite{Verheijen2001} and is briefly described here. The 3D mask is put at eight different positions in the noise-added mock datacube which are devoid of \HI\ line emission and summing the flux along the frequency axis provides 8 signal free maps. A noise map is then created by calculating for each pixel in this map the variance of the 8 values of the corresponding pixels in the signal free maps. Finally, a S/N map is acquired by dividing the $N_\mathrm{HI}$ map by the noise map. For the 120 noise-added mock datacubes of each galaxy, we create a corresponding 3D mask, a column density map, and a S/N map. We also measure morphological asymmetries using Eq. \ref{AsyMod} including pixels above a column density threshold under consideration.
\par To create global profiles with noise, we apply the same 3D mask used to create the column density map and sum the flux in each channel of the mask-applied datacube. Asymmetries in the global profile are measured using Eq. \ref{Aflux}. For each channel of the profile, the uncertainty in the flux is estimated by multiplying the noise with the square root of the number of independent beams enclosed within the mask. The uncertainty in the flux varies with channel as the 3D mask has different number of pixels in each channel. We then calculate the S/N value for each channel and the S/N value of the global profile is defined as the average of the S/N values in all the channels of the profile. Note that this is not the maximum S/N of the total flux when using the matched filter technique described in \cite{Saintonge2007}. Since we are interested in measuring profile shapes at a fixed velocity resolution, we prefer to express the S/N at a fixed velocity resolution as well.
\par In Appendix \ref{appendix:A}, we present the properties of mock galaxies mentioned in this section along with the catalogued GroupNumber of the FoF halo in the Recal25 run of the \textsc{eagle} simulations. For each galaxy, we present noise-free column density maps at various angular resolutions and a noise-free global profile in the form of an \HI\ atlas in Appendix \ref{appendix:B}.
\section{Estimating bias in morphological asymmetry index due to observational constraints}\label{sec:Moprhasymmetry}
\par Characterization of the asymmetry in the \HI\ morphology of galaxies is prone to uncertainties due to various observational parameters such as angular resolution, column density threshold, and the S/N ratio of this threshold. To characterise morphological asymmetries in the outer \HI\ disc of galaxies a low column density threshold is required and to be able to detect gas in the outer \HI\ disc, a high sensitivity to low surface density emission is imperative. This can be achieved by degrading the angular resolution of observations or, when possible, by a extending the integration period. Lowering the resolution reduces the noise in units of Kelvin and thereby improves the S/N ratio corresponding to a certain column density level in a column density map. However, the resolution should not be lowered so much that asymmetric features in \HI\ maps of galaxies are washed out. In addition, the chosen column density threshold should be unaffected by noise to obtain a robust measure of the morphological asymmetry. In this section, we quantify the constraints on column density threshold, resolution and S/N that are required to reliably measure the morphological asymmetry  of galaxies using the \Amod\ index.  
\subsection{The effect of column density threshold on \Amod.}\label{CDonAmod}
\par
\begin{figure*}
    \centering
    \includegraphics[width=0.8\textwidth]{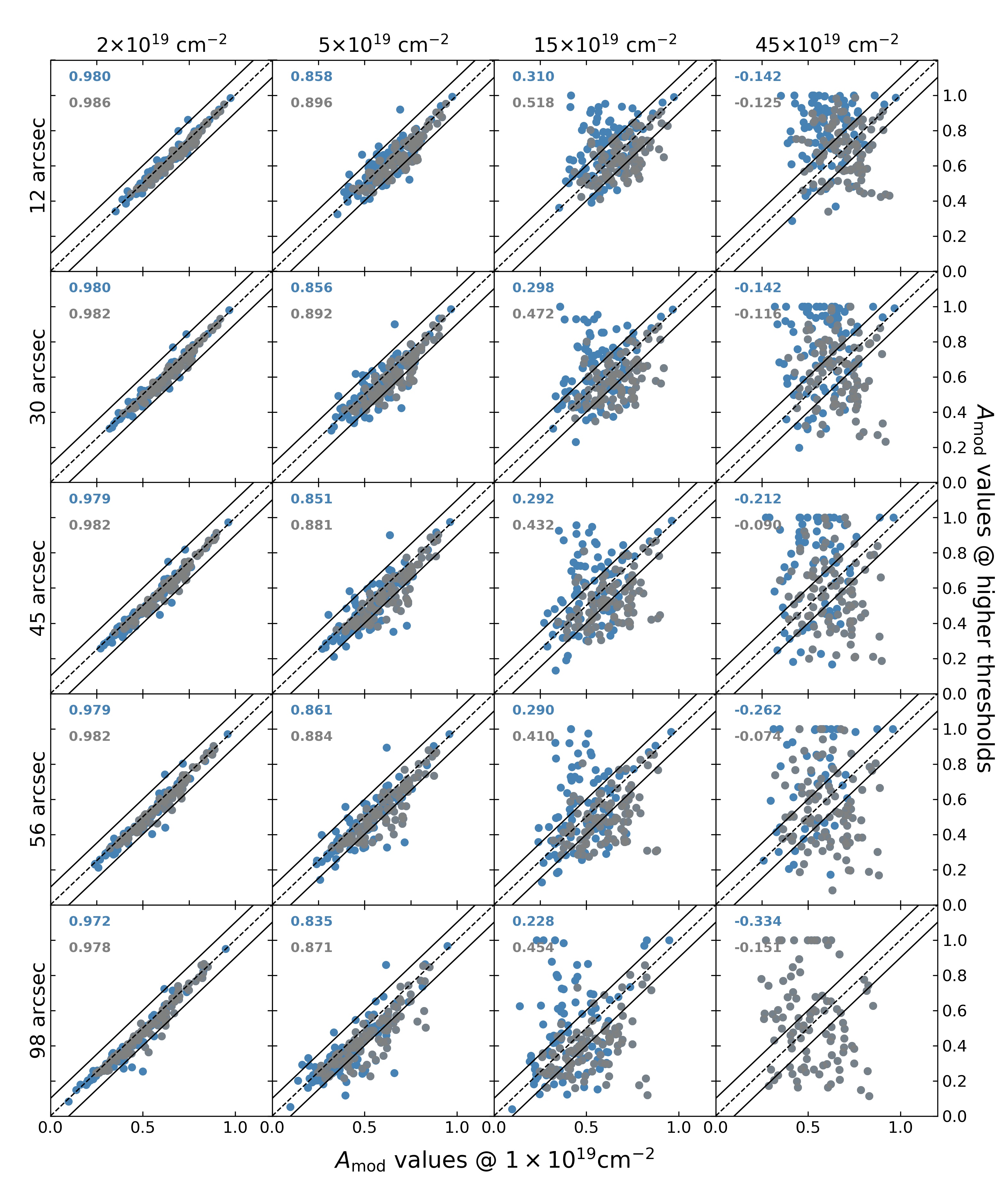}
    \caption{In this figure, asymmetries measured in noise-free mock column density maps are shown. We compare asymmetry values measured at a reference column density threshold of $1\times 10^{19}$\cm\ to those measured at higher thresholds of $2,5,15,45\times10^{19}$\cm. This comparison is performed at different resolutions of the mock maps. Blue symbols indicate the asymmetry values of the whole sample while grey symbols indicate a small subset of mock galaxies that have high column density gas above $45\times10^{19}$\cm\ at the lowest resolution of 98 arcsec$^{2}$. In each panel, the numbers in the top left corner indicate the Spearman correlation coefficient between the \Amod\ values being compared and colors indicate the sample under consideration. The dashed black line indicates the line of equality and solid black lines indicate 10 per cent deviation from equality.}
    \label{fig:CompareScatter}
\end{figure*}

 In order to identify an acceptable low value of the column density threshold that can characterise asymmetries in the outer \HI\ disc of galaxies, we take advantage of the noise-free column density maps derived from the \textsc{eagle} simulations, in which the column densities can be reliably traced to values as low as $1\times10^{17}$\cm. Observationally, however, this is difficult to achieve. The HALOGAS survey \citep{Halogas2011} is one of the deepest surveys of nearby galaxies and for the four galaxies in their pilot survey observed with the WSRT, an integration period of 120 hours was required to reach $3\sigma$ column density sensitivity of about $1\times10^{19}$\cm. To observe gas in galaxies at column densities lower than this while still having sufficient angular resolution, we would need longer integration periods, which is difficult to achieve. Therefore, we use the column density threshold of $1\times10^{19}$\cm\ as a reference threshold. We compare the \Amod\ values of the \textsc{eagle} galaxies measured at this reference threshold to the \Amod\ values at higher thresholds of 2,  5,  15, and 45 $\times10^{19}$\cm. We measure the \Amod\ values at the above-mentioned thresholds in column density maps at resolutions of $12\times17$, 30, 45, 56, and 98 arcsec. Thereafter, at each resolution we calculate the Spearman correlation coefficient between the \Amod\ values at the reference threshold and the \Amod\ values at higher thresholds. The comparison of \Amod\ values along with the correlation coefficients are illustrated in figure \ref{fig:CompareScatter}. 
\par
\begin{figure}
    \centering
    \includegraphics[width=0.5\textwidth]{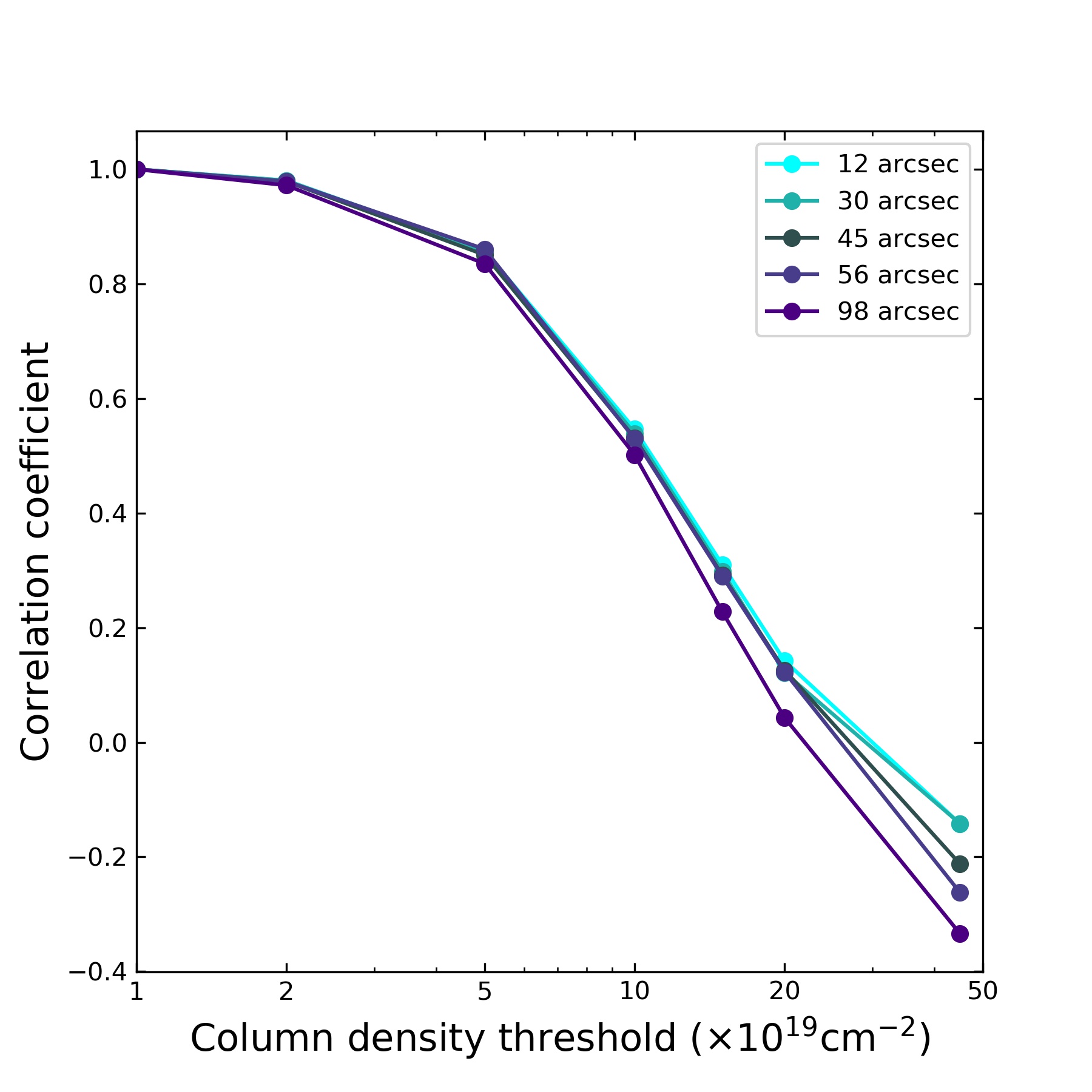}
    \caption{Correlation coefficients of the full sample are plotted as a function of the column density threshold. The coefficients correspond to those noted in the various panels of Fig \ref{fig:CompareScatter}. Additionally, correlation coefficients at $10\ \text{and}\ 20\times10^{19}$\cm\ are also included here.}
    \label{CompareCorrelation}
\end{figure}

 From Fig. \ref{fig:CompareScatter}, we find that irrespective of the angular resolution most of the galaxies that have \Amod\ values at and below a column density threshold of $5\times10^{19}$\cm\ are within 10 per cent of the reference \Amod\ value. This also reflects in the value of the Spearman correlation coefficients shown in figure \ref{fig:CompareScatter} and thus we conclude that \Amod\ values at and below $5\times10^{19}$\cm\ are well correlated to the \Amod\ values at the reference threshold. Figure \ref{CompareCorrelation} illustrates this as well: the correlation coefficient drops quickly when comparing \Amod\ values above $5\times10^{19}$\cm. This implies that the asymmetries at the reference threshold of $1\times10^{19}$\cm\ are similar to asymmetries at thresholds up to $5\times10^{19}$\cm. Thus, to investigate the effect of environmental processes on outer parts of \HI\ disc of galaxies, \Amod\ values should be measured at or below column density thresholds of $5\times10^{19}$\cm. If observational limitations do not allow this and higher column density thresholds have to be applied while measuring \Amod\ then it is important to note that the \Amod\ values may not reflect the asymmetries in the outer parts of the \HI\ disc.
\par Due to beam dilution, at the lowest angular resolution of 98 arcsec (or 8 kpc at the adopted distance of 17 Mpc), high column density clumps above $45\times10^{19}$\cm\ are not present in many of the noise-free column density maps. In our sample of $189$ mock galaxies, 90 galaxies do not have gas above this threshold at 98 arcsec resolution and are therefore not shown in the bottom right panel of Fig. \ref{fig:CompareScatter}. Therefore, we define a subsample of mock galaxies for which we can measure the \Amod\ values at 98 arcsec resolution and at a threshold of $45\times10^{19}$\cm. This subsample of mock galaxies is shown in grey in Fig. \ref{fig:CompareScatter}. We show the correlation coefficients of the complete mock sample as well as for the subsample in each panel of Fig. \ref{fig:CompareScatter}.

\subsection{The effect of resolution on \Amod}\label{ResonAmod}
\par In the previous subsection, we investigated the effect of changing the column density threshold on the \Amod\ values. In this subsection, we examine how changing the resolution affects the \Amod\ values. We again use noise-free column density maps of the \textsc{eagle} mock galaxies to ensure that the asymmetry values are not being affected by noise. We use column density maps at angular resolutions of $12\times17$, 30, 45, 56, and 98 arcsec$^{2}$, and an additional resolution of 120 arcsec$^{2}$ to further reduce the number of resolution elements across the \HI\ map. At each angular resolution, we have measured the \Amod\ values at a column density threshold of $5\times10^{19}$\cm\ as this is the upper limit on the column density threshold established in Section \ref{CDonAmod}. The column density maps of the mock galaxies above the threshold of $5\times10^{19}$\cm\ may be rather irregular. Therefore, we identify the minimum-sized rectangular box that encompasses all pixels above this column density threshold and calculate the number of beams that fit across the diagonal of this box. 
\par
\begin{figure*}
    \centering
    \includegraphics[width=\textwidth]{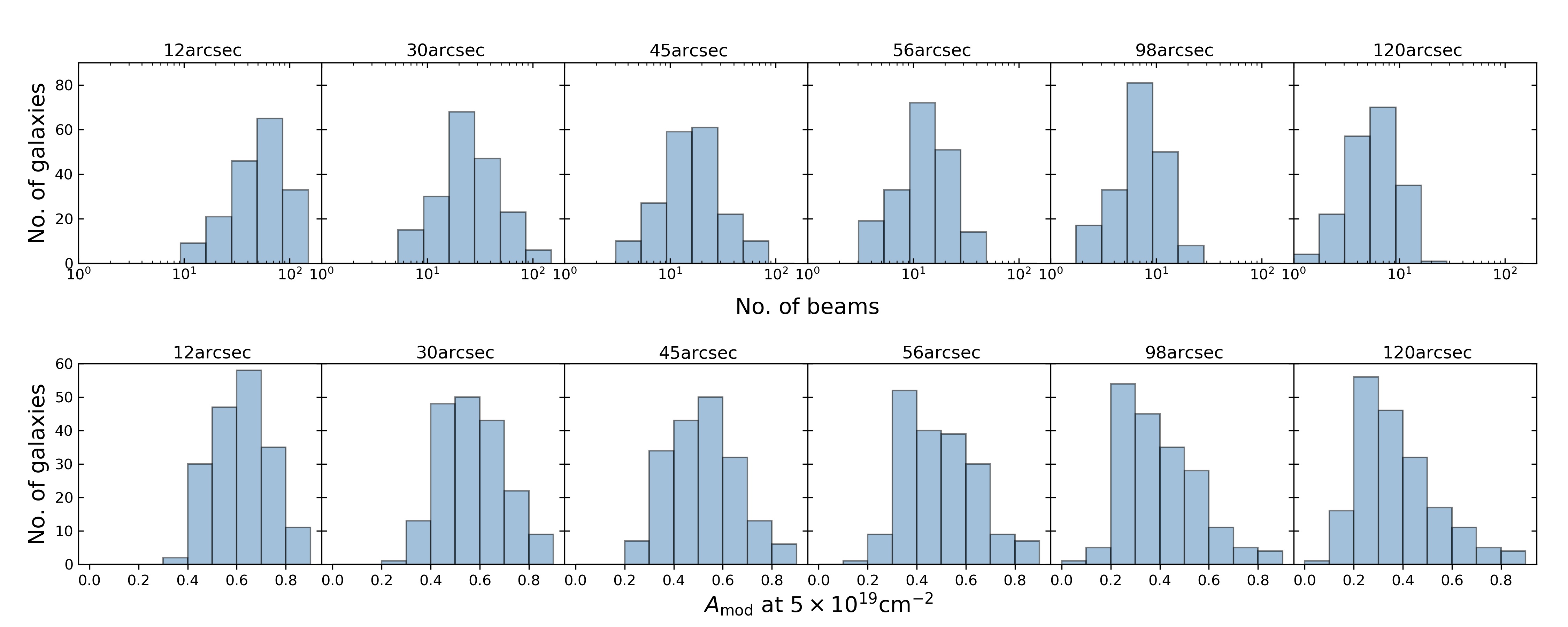}
    \caption{The top panels of this figure shows the number of beams across galaxies at different angular resolutions. The sizes are measured in noise-free column density maps of mock galaxies with a column density threshold of $5\times10^{19}$\cm. The bottom panels show the asymmetry values measured in noise-free column density maps with the same threshold.}
    \label{fig:SizeAmod@5}
\end{figure*}

 The upper panels of Fig. \ref{fig:SizeAmod@5} illustrate the distribution of the number of beams across the diagonal at different angular resolutions in the noise-free maps of the mock galaxies. At the highest resolution of $12\,\mathrm{arcsec} \times 17\,\mathrm{arcsec}$, all the galaxies are well resolved. As the angular resolution is lowered, the number of beams across a galaxy reduces. The lower panels of Fig. \ref{fig:SizeAmod@5} illustrate the distribution of \Amod\ values of the mock galaxies at different angular resolutions. At the highest angular resolution, all the galaxies are well resolved and as the angular resolution is lowered galaxies become progressively unresolved. Similarly, the \Amod\ distribution shifts to lower asymmetry values as the resolution is lowered. At the lowest resolution, the bulk of the distribution has \Amod$\,<0.5$ with a tail at high asymmetry values.
\par
\begin{figure*}
    \centering
    \includegraphics[width=\textwidth, trim={0cm 2cm 0cm 2cm}, clip]{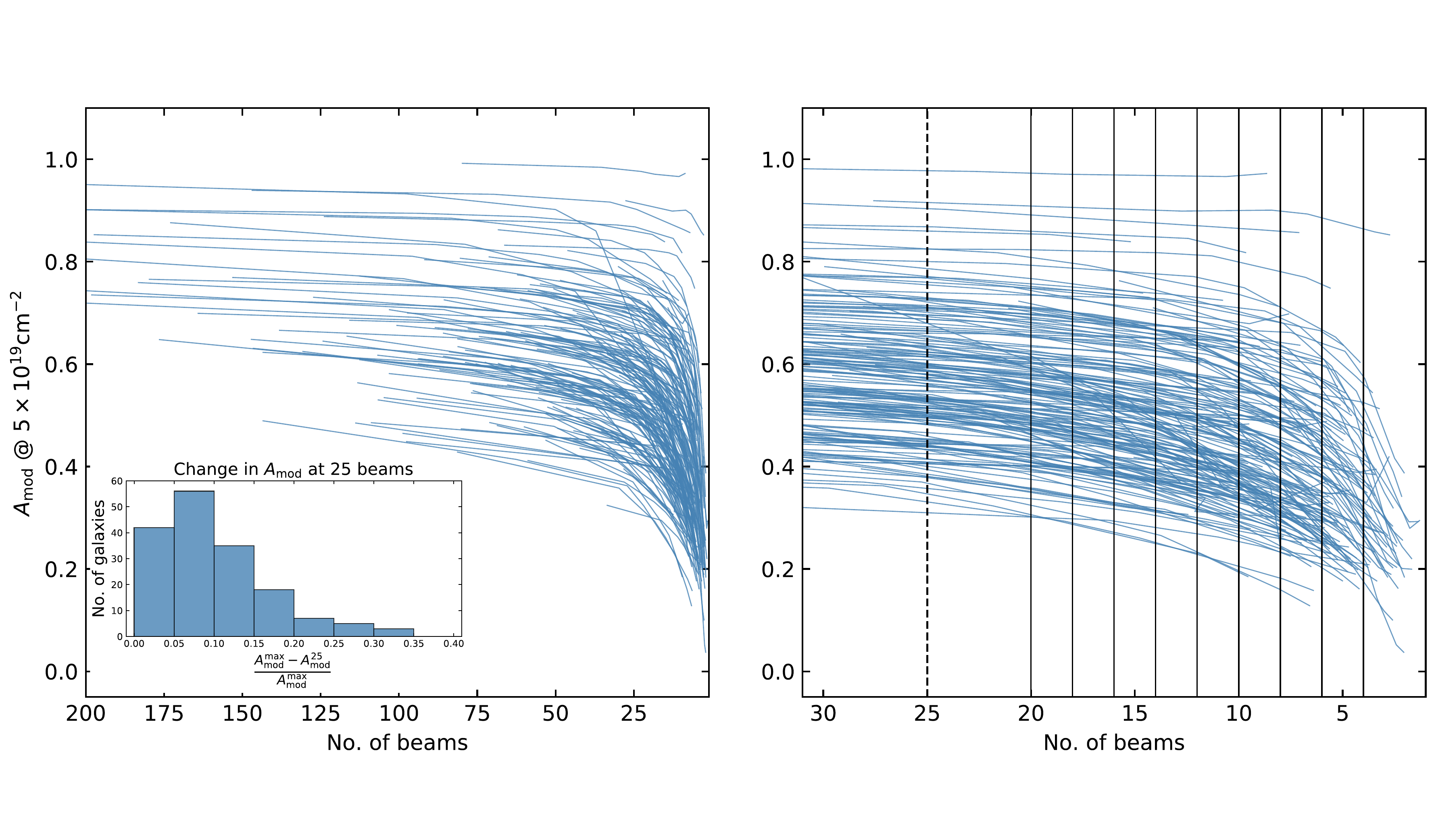}
    \caption{This figure shows how asymmetry values change as the number of beams across a galaxy is lowered. The asymmetry is measured with a threshold of $5\times10^{19}$\cm. In the panel on the left, full range of sizes measured in beams is shown, while in the right panel, sizes in the range of 0 to 30 beams is illustrated. For each galaxy in our sample, we measure the relative change in \Amod\ value at 25 beams with respect to the maximum \Amod\ value for the galaxy. The histogram of these values is shown in the left panel.}
    \label{fig:CompareResolution}
\end{figure*}

\begin{figure*}
    \centering
    \includegraphics[width=0.9\textwidth]{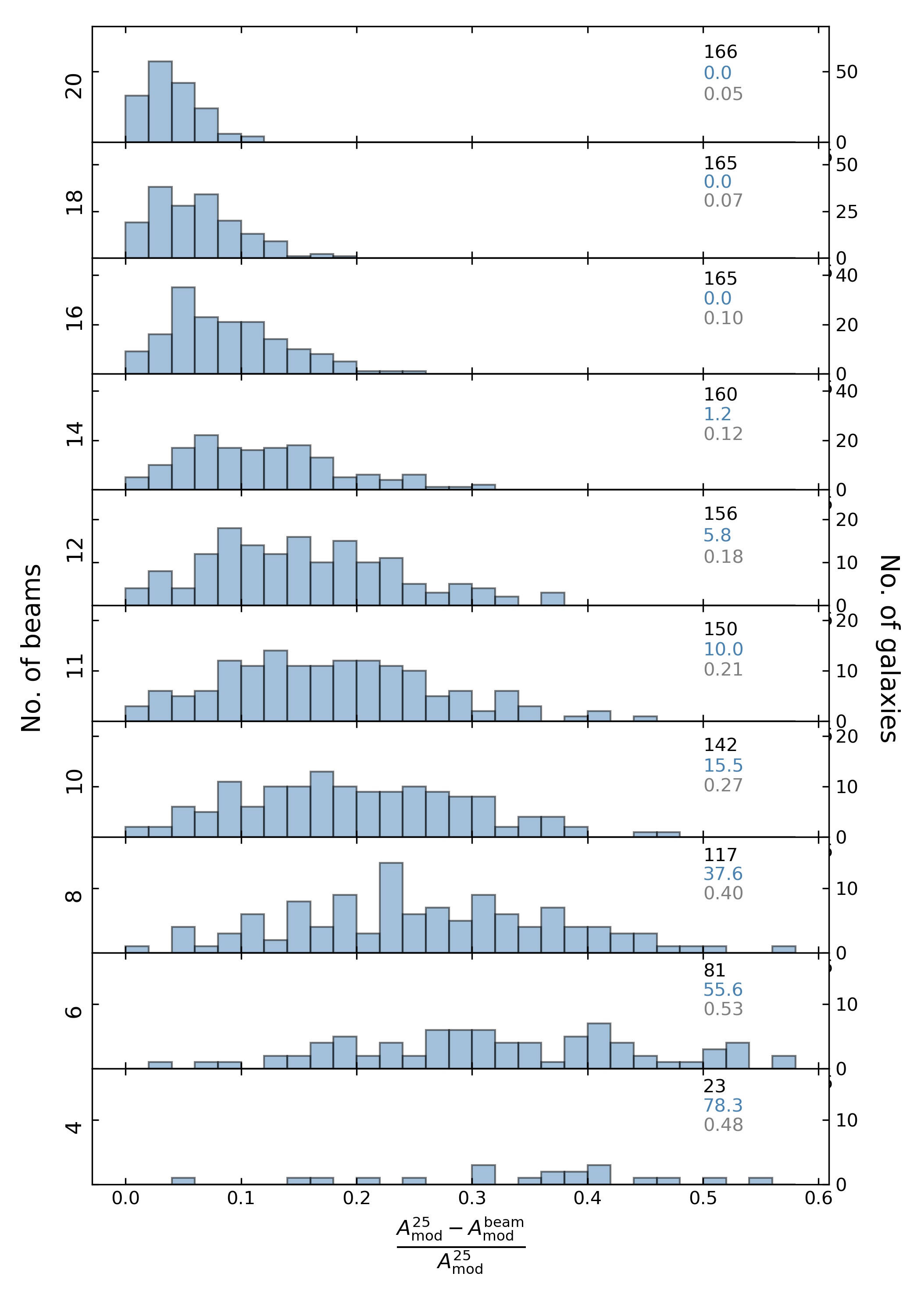}
    \caption{This figure illustrates how asymmetry values change at different number of resolution elements across a galaxy, with respect to a reference asymmetry value at 25 beams. The asymmetry values are measured at a column density threshold of $5\times10^{19}$\cm. The number of galaxies in each panel is mentioned in black in the top right corner, in blue we indicate the percentage of the sample in each panel that has more than a 30 per cent change in \Amod\ values and in grey we mention the median of the histogram in each panel.}
    \label{fig:ChangeAmodResolution}
\end{figure*}

 In an absolute sense, the number of beams across a column density map depends not only on the angular resolution of the column density map but also on the distance to the galaxy. In order to resolve nearby galaxies a large synthesised beam may be adequate however this may not be sufficient to resolve galaxies that are farther away. Therefore, we define resolution in terms of the number of beams across a galaxy and aim to quantify the minimum required number of beams to reliably measure \Amod\ values. Lines in the left panel of Fig. \ref{fig:CompareResolution} illustrate how \Amod\ values for an individual galaxy vary as a function of the number of beams across that galaxy. From Fig. \ref{fig:CompareResolution} it is evident that the \Amod\ value changes little when the number of beams across a galaxy exceeds 25 while the \Amod\ value changes rapidly when the number of beams is further reduced below 25. The histogram shown at the bottom left of the left panel of Fig. \ref{fig:CompareResolution} illustrates the relative change in the \Amod\ value at 25 beams with respect to the \Amod\ value at the maximum number of beams. We find that 20 per cent of the sample shows more than 15 per cent relative change in the \Amod\ value. To quantify the change in \Amod\ with resolution, we choose the \Amod\ value at 25 beams as our reference and measure the relative change in \Amod\ with respect to our reference as the number of beams across a galaxy decreases. The right panel of Fig. \ref{fig:CompareResolution} shows the variation in \Amod\ values below 25 beams in more detail while Fig. \ref{fig:ChangeAmodResolution} shows histograms of the relative change in \Amod\ at different number of beams. 
\par In the top right corner of each panel in Fig. \ref{fig:ChangeAmodResolution}, we indicate in black the number of galaxies that are shown in the histogram, in blue we indicate the percentage of the sample in that panel that shows more than a 30 per cent change in \Amod\ and in grey the median value of relative change in \Amod\ . For example, at 20 beams none of the galaxies in the sample have more than a 30 per cent change in \Amod\ with respect to the reference \Amod\ value at 25 beams while at 12 beams, about 6 per cent of the sample shows more than 30 per cent change in \Amod\ . From Fig. \ref{fig:ChangeAmodResolution}, we find that as the resolution reduces, a larger fraction of the sample shows more than 30 per cent change in \Amod\ . We find it acceptable to have 10 per cent of our sample showing more than 30 per cent change in \Amod\ value, which from Fig. \ref{fig:ChangeAmodResolution} occurs at 11 beams. Therefore, we conclude that at least 11 beams across a galaxy are required to prevent a significant systematic reduction in the \Amod\ value due to resolution. We recommend the reader to choose the highest possible resolution in number of beams that their observational limitations allow for while recognizing that lowering the resolution may result in \Amod\ values that are lower than the intrinsic \Amod\ value.  
\subsection{Effect of signal-to-noise on asymmetry}\label{sec:SNonAmod}
\par In an actual observed datacube with noise, the angular resolution, the column density threshold and the S/N value associated with that threshold are interlinked. To investigate the minimum required S/N value that provides robust asymmetry measurements, we measure \Amod\ values of mock galaxies with and without added noise, at a threshold of $5\times10^{19}$\cm\ and at angular resolutions of $12\times17$, 30, 45, 56 arcsec$^{2}$. Additionally, we only include galaxies from the mock sample that are at least 11 beams across at the various angular resolutions. By excluding galaxies with fewer than 11 beams across, we ensure that the change in the morphological asymmetry in the noise-added mock cubes would result solely from the addition of noise. As described in Section \ref{MockHICubes}, noise is added to the mock datacubes such that the threshold value of $5\times10^{19}$\cm\ has S/N ratios of 1, 2, 3, 4, and 5. We recall that adding noise not only affects the calculation of \Amod\ but also the shape of the mask that is applied when constructing the column density map.
\par
\begin{figure*}
    \centering
    \includegraphics[width=0.95\textwidth]{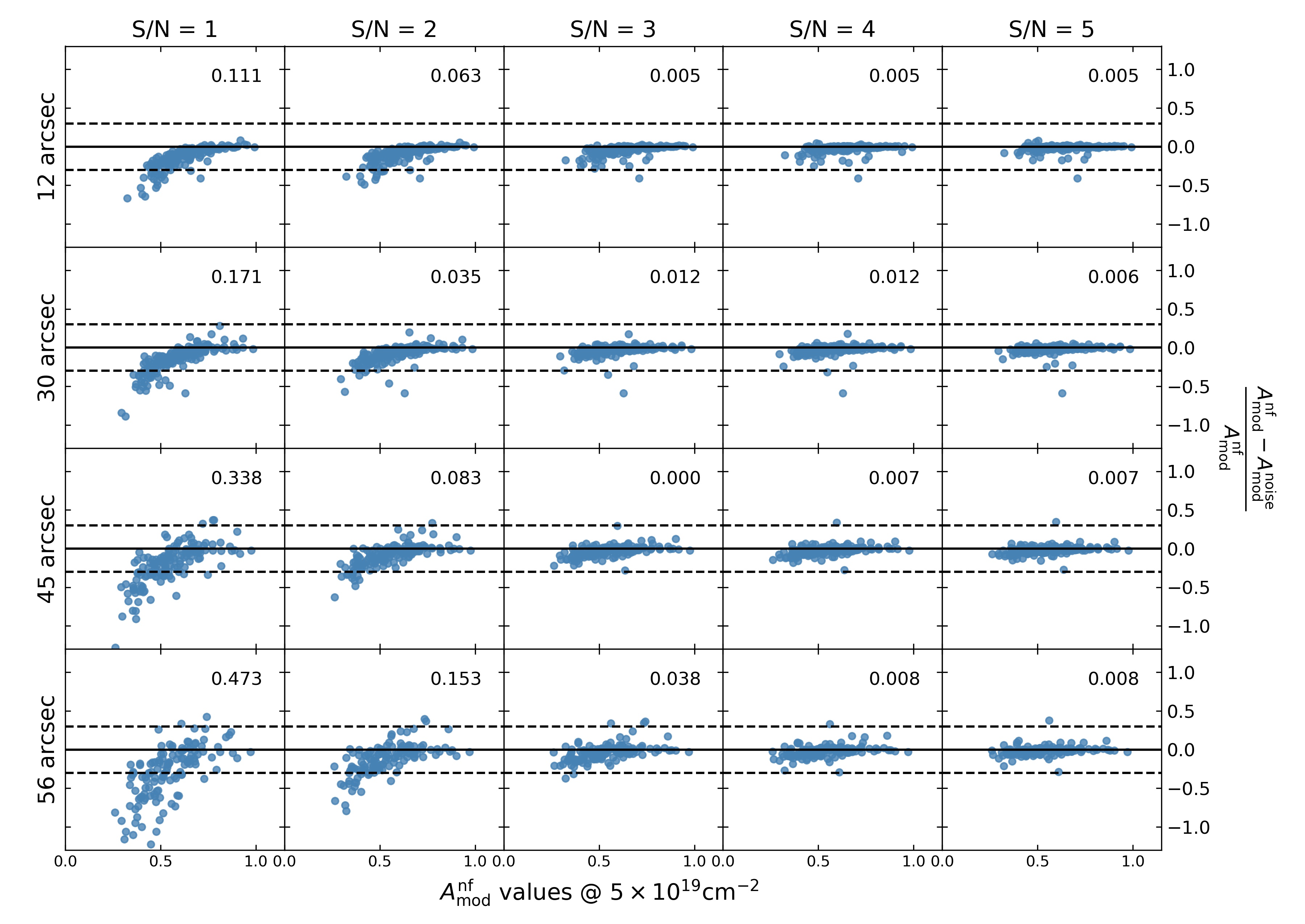}
    \caption{Relative change in asymmetry values measured at $5\times10^{19}$\cm\, at different signal-to-noise ratios. The amount of noise added to the mock \HI\ cubes reduces from left to right in each row as the signal-to-noise improves. In each panel, dashed black lines indicate $\pm30$ per cent change in asymmetry value and the numbers at the top right corner indicate the fraction of the sample with more than a 30 per cent change.}
    \label{fig:CompareSN}
\end{figure*}

 For every mock galaxy with added noise, we measure the relative change in \Amod\ with respect to \Amod\ measured from the noise free column density maps as the S/N ratio varies. By implementing constraints on the column density threshold and the resolution, we aim to minimise the effect of these parameters on the measured \Amod\ values. In Fig. \ref{fig:CompareSN} we illustrate the relative change in \Amod\ at different S/N values for the column density threshold of $5\times10^{19}$\cm\ and calculate the fraction of the galaxies in the sample that show more than $30$ per cent change in their \Amod\ value. This fraction is shown in the top right corner of each panel. From Fig. \ref{fig:CompareSN} we find that the fraction of galaxies in the sample for which \Amod\ changes more than $30$ per cent reduces as the signal-to-noise increases from S/N$=$1 to S/N$=$5. When S/N$\,>$3, the effect of noise on the intrinsic \Amod\ values is minimal, i.e. less than 5 per cent of the galaxies in the sample have their \Amod\ value change by more than 30 per cent. Thus, we recommend that the chosen column density threshold should have a minimum S/N of 3 to prevent a significant systematic increase in the measured \Amod\ value due to noise.
\par Note that galaxies that are intrinsically very asymmetric (\Amod$>0.75$), the relative change in \Amod is small due to the fact that \Amod values cannot increase above 1 after the addition of noise.

\section{Asymmetries in the global profiles}\label{sec:GPasymmetry1}
\begin{figure}
    \centering
    \includegraphics[width=0.5\textwidth]{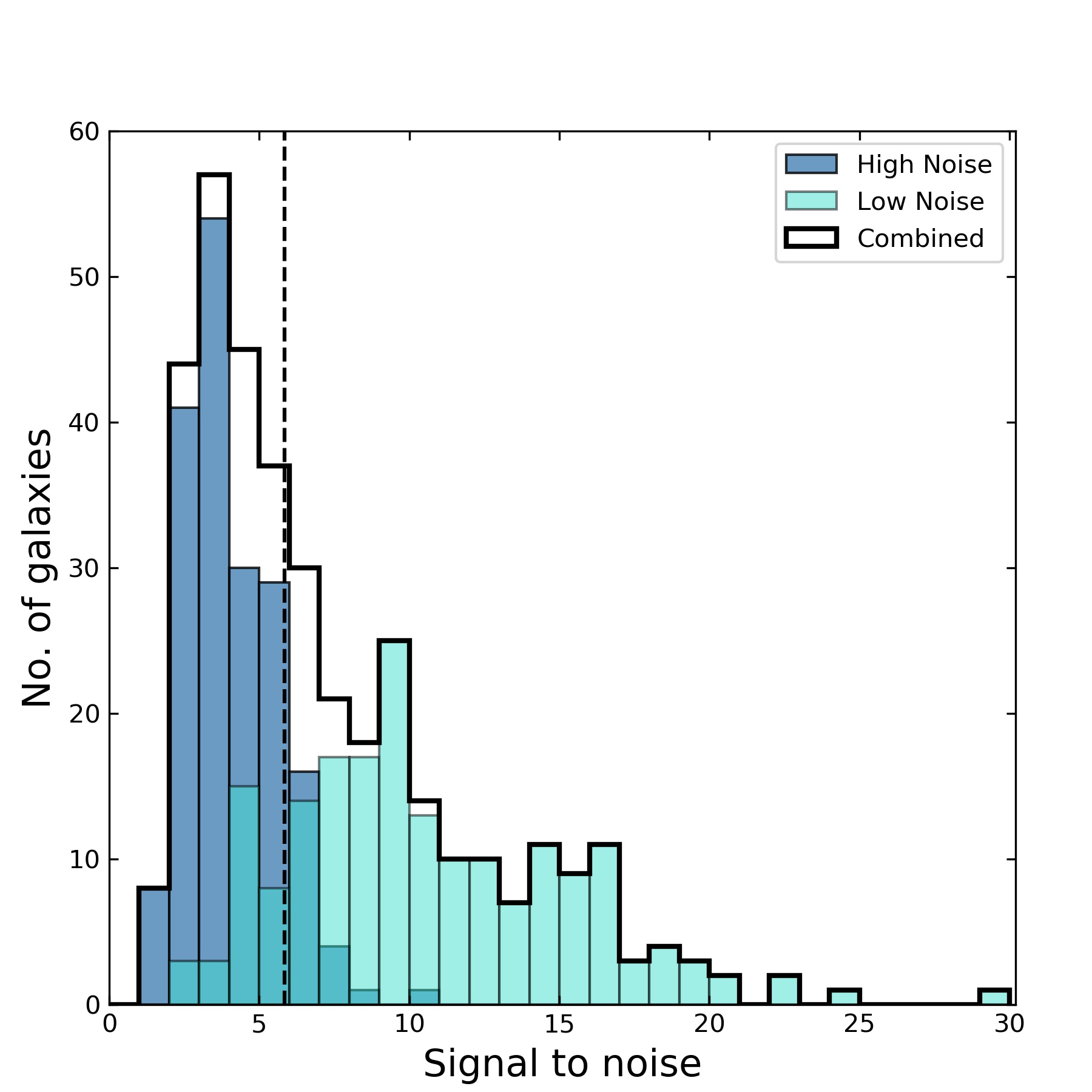}
    \caption{Distribution of S/N values of the two sets of global profiles at 56 arcsec$^{2}$ and $20\,\mathrm{km}\,\mathrm{s}^{-1}$. The distribution of the combined sample is also shown with a solid black line. We divide the combined sample into roughly equal halves by splitting the sample at S/N$=5.84$. This S/N value is shown with a dashed black line.}
    \label{fig:SN_hist}
\end{figure}

In the previous section, we concluded that a column density threshold of $5\times10^{19}$\cm\ or lower, at least 11 beams across the galaxy and a S/N of 3 or higher are required to obtain a robust estimation of \Amod\ values. In this section, we investigate the effect of adding noise to the global profiles on measured \Aflux\ values. For this, we use the noise-added mock datacubes described in Section \ref{sec:SNonAmod} at an angular resolution of 56 arcsec$^{2}$ and a velocity resolution of $20\,\mathrm{km}\,\mathrm{s}^{-1}$ to which we add noise such that the column density threshold of $5\times10^{19}$\cm\ has S/N=3 in the integrated column density map. As described in Section \ref{sec:MockSample}, we create global profiles with noise by first applying a mask to the mock datacube before adding the flux in each channel. We also characterise the S/N value of the global profile by averaging the S/N in each channel. Fig. \ref{fig:SN_hist} shows the distribution of S/N values, indicated by the label `low noise', where we find that very few galaxies have a global profile with S/N$\,<5$. It is in global profiles with low S/N values, however, where the effect of noise on the global profile asymmetries would be evident. We therefore create another set of noise-added global profiles where we add noise to the mock datacubes such that the column density threshold of $5\times10^{19}$\cm\ has a S/N=1 in the integrated column density map. In Fig. \ref{fig:SN_hist}, we see that the S/N histogram of the second set, labeled as `high noise', fills the gap in low S/N values left by the first set of global profiles. We combine the two sets of noise-added global profiles to study the effect of noise on the \Aflux\ values which are further divided into roughly equal halves by splitting the sample at S/N$=5.84$. The global profiles subset with S/N $\leq5.84$ is called the `Low S/N' subset while that with S/N $>5.84$ is called the `High S/N' subset. To create noise-free global profiles we use the mock datacubes without noise at a resolution of 56 arcsec and $20\,\mathrm{km}\,\mathrm{s}^{-1}$ where we do not apply a 3D mask but include the flux in all the pixels in each channel of the datacube.
\par
\begin{figure*}
    \centering
    \includegraphics[width=0.98\textwidth]{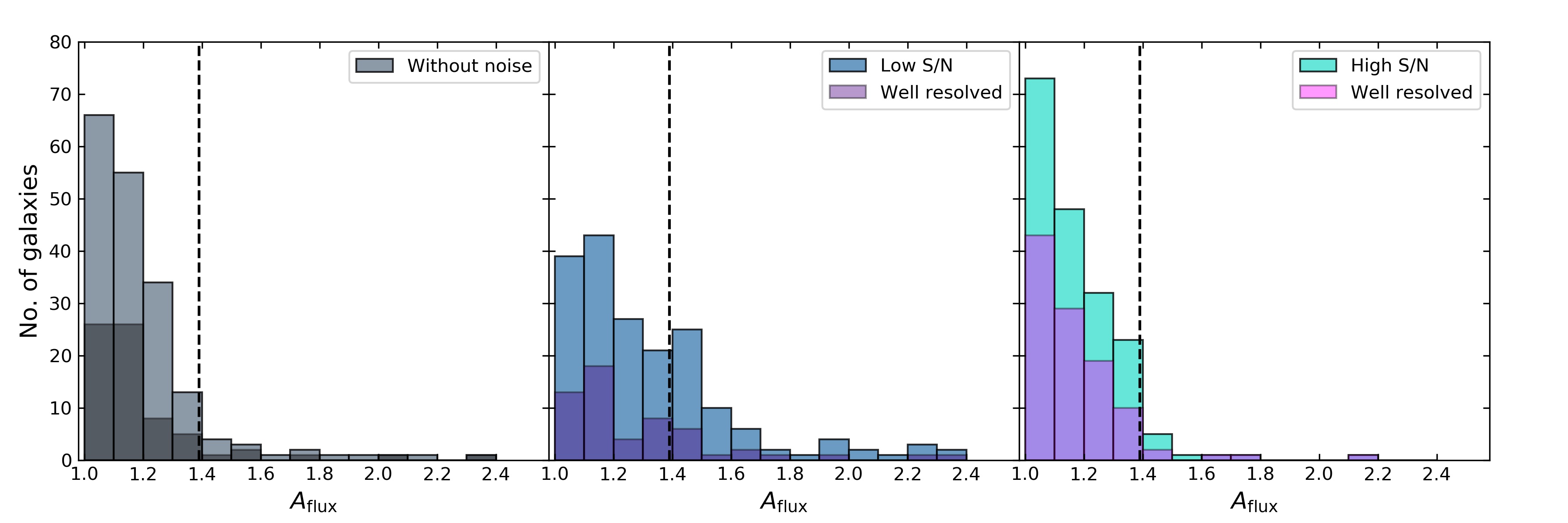}
    \caption{The left panel shows the distribution of \Aflux\ values without noise whereas the middle and right panel show the distribution of \Aflux\ for the high and low S/N subsets of noise-added global profiles. In each panel, we indicate the \Aflux\ values of global profiles that have more than 10 resolution elements across the 20 per cent width. The global profiles were created from datacubes at an angular resolution of 56 arcsec$^{2}$ and velocity resolution of $20\,\mathrm{km}\,\mathrm{s}^{-1}$. The vertical dashed line shows the $3\sigma$ asymmetry threshold of the AMIGA sample.}
    \label{fig:AfluxHist}
\end{figure*}

 Using Eq. \ref{Aflux}, we measure \Aflux\ values in global profiles without noise and the two subsets of noise-added global profiles. We illustrate these \Aflux\ values as histogram distributions in Fig. \ref{fig:AfluxHist}. \cite{Espada2011} find that the distribution of \Aflux\ values of the refined AMIGA sample can be parameterised with a half-Gaussian distribution that has a $\sigma=0.13$. This dispersion of the half-Gaussian sets an upper limit on the intrinsic dispersion of the isolated galaxies from the AMIGA sample. \cite{Espada2011} define asymmetric profiles as those having \Aflux$\,>3\sigma$ (i.e. \Aflux$\,>1.39$) and they find that 2 per cent of galaxies in their sample satisfy this criterion. Using the same criterion as that of \cite{Espada2011} as a reference, we find that 9.8 per cent of the sample has \Aflux$\,>1.39$ in global profiles without noise. In their analysis, \cite{Espada2011} use well-resolved global profiles that have more than 10 resolution elements across the 20 percent profile width. For the sake of a fair comparison, we also measure the rate of asymmetry in global profiles with more than 10 resolution elements of the noise-free sample. We find that 9.7 per cent of the noise-free global profiles have \Aflux$\,>1.39$. This indicates that noise-free global profiles of mock galaxies are more asymmetric than the isolated galaxies described in \cite{Espada2011}, which is expected because galaxies that form the mock sample are not selected from particularly isolated environments and therefore may have asymmetries due to environmental effects. In addition, the simulated galaxies may intrinsically be more asymmetric than real galaxies as shown by \cite{Bahe2016}. They find that a majority of their sample has vertically disturbed \HI\ discs. \cite{Bahe2016} also find that more than 80 per cent of their sample has \HI\ holes larger than those seen in observed galaxies, which result from the feedback implementation in \textsc{eagle} simulations.

\par In global profiles of the high and low S/N subsets, we find that 7.0 per cent and 32.0 per cent of the subsets have asymmetric profiles respectively. Considering global profiles with more than 10 resolution elements, we find that 5.6 per cent and 23.2 per cent of the high S/N and low S/N subset of the noise-added global profiles have \Aflux$\,>1.39$. \cite{Watts2020} demonstrated that an intrinsically symmetric global profile may show noise-induced asymmetries after the addition of noise, the degree of which is dependent on the S/N of the profile. They did not investigate, however, how the addition of noise affects the asymmetry of intrinsically asymmetric profiles. \cite{Yu2020} also find that in the low S/N regime, fractional uncertainties in the \Aflux\ index increases as the S/N decreases. When compared to the noise-free sample, the rate of asymmetry in the high S/N sample has reduced as asymmetric regions of low column density may have been excluded when the 3D mask is applied. Similarly, the rate of asymmetry in the low S/N sample has increased mainly due to the addition of noise. It should be noted, however, that choosing a fixed threshold of \Aflux\ (i.e. independent of S/N) to decide the fraction of asymmetric galaxies in a sample would introduce a positive bias in the low S/N bins \citep{Watts2020}. This may be especially relevant in the low S/N subset. In the following paragraphs we investigate in more detail the cause for the different rates of asymmetries in the high and the low S/N subsets.
\par
\begin{figure}
    \centering
    \includegraphics[width=0.5\textwidth]{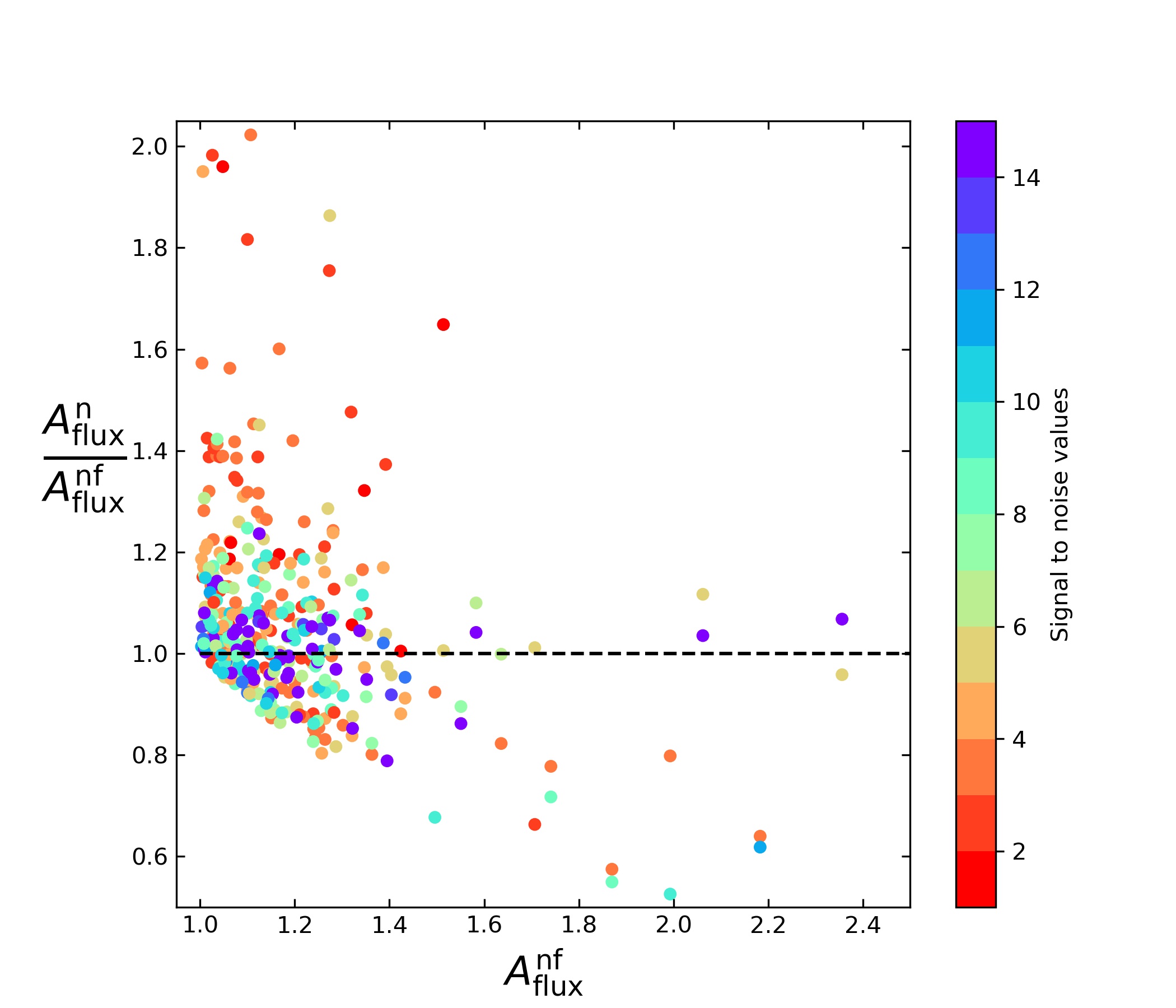}
    \caption{This figure illustrates the ratio of the noise-free \Aflux\ to the noise-added \Aflux\ with respect to the noise-free \Aflux\ . Colours indicate the S/N of the noise-added global profile. For galaxies with a symmetric noise-free global profile, the addition of noise induces asymmetry in them, while for galaxies with intrinsically asymmetric profiles the noise-added asymmetry may increase or decrease.}
    \label{fig:AfluxCompare}
\end{figure}

 To verify how the noise affects the global profile asymmetries, we examine how the ratio of the \Aflux\ values in the two sets of noise-added global profiles to those without noise varies with respect to the \Aflux\ values without noise. This is illustrated in Fig. \ref{fig:AfluxCompare}. In this figure, the S/N value of the global profile is indicated with colours. Fig. \ref{fig:AfluxCompare} shows that for galaxies with an intrinsically symmetric global profile, the \Aflux\ values increase upon addition of noise. This is similar to the effect observed by \cite{Watts2020} in their model global profiles. On the other hand, for galaxies with intrinsic asymmetries in their profile, the addition of noise may increase or decrease the \Aflux\ values. We divide the sample into 9 bins in S/N value with 40 global profiles in each bin and an additional high S/N bin with 13 global profiles. In each bin we measure the median S/N, the median of the ratio of the noise-added \Aflux\ value to the noise-free \Aflux\ value (hereafter \Aflux\ ratio) as well as the $75^\mathrm{th}$ and $25^\mathrm{th}$ percentile of the \Aflux\ ratio. These values are presented in Table \ref{tab:GPAsymValues1} and Fig. \ref{fig:AfluxCompare}, from which we conclude that global profiles with S/N$\,<5.62$ have a high median \Aflux\ ratio as well as a large spread in their inter-quartile range of the \Aflux\ ratio. This indicates that the noise-added \Aflux\ values with S/N$\,<5.62$ are not representative of the actual noise-free \Aflux\ as the added noise most often increases the asymmetries in the global profiles but could occasionally decrease the asymmetry as well. Therefore we recommend that global profiles with S/N$\,>5.5$, which is the lower bound of the S/N$\,=5.62$ bin, for robust measurement of global profile asymmetries.
\begin{table}
    \centering
    \caption{Median S/N and \Aflux\ ratio of galaxies in different bins of S/N values as shown in Fig. \ref{fig:AfluxCompare}. We divide the sample in 9 bins with 40 global profiles each, while the last bin contains 13 global profiles.}
    \begin{tabular}{c c c c}
    \hline
    Median S/N & Median \Aflux\ ratio & $75^\mathrm{th}$ percentile & $25^\mathrm{th}$ percentile\\
    \hline
    2.23 & 1.16 & 1.35 & 1.01\\
    3.13 & 1.10 & 1.39 & 1.00\\
    3.84 & 1.05 & 1.17 & 0.92\\
    4.60 & 1.07 & 1.17 & 0.96\\
    5.62 & 1.00 & 1.07 & 0.95\\
    6.88 & 1.00 & 1.11 & 0.94\\
    8.98 & 1.03 & 1.10 & 0.93\\
    11.24 & 1.00 & 1.05 & 0.95\\
    15.15 & 1.01 & 1.06 & 0.96\\
    19.71 & 1.01 & 1.04 & 0.99\\
    \hline
    \end{tabular}
    \label{tab:GPAsymValues1}
\end{table}
\par
\begin{figure*}
    \centering
    \includegraphics[width=0.98\textwidth]{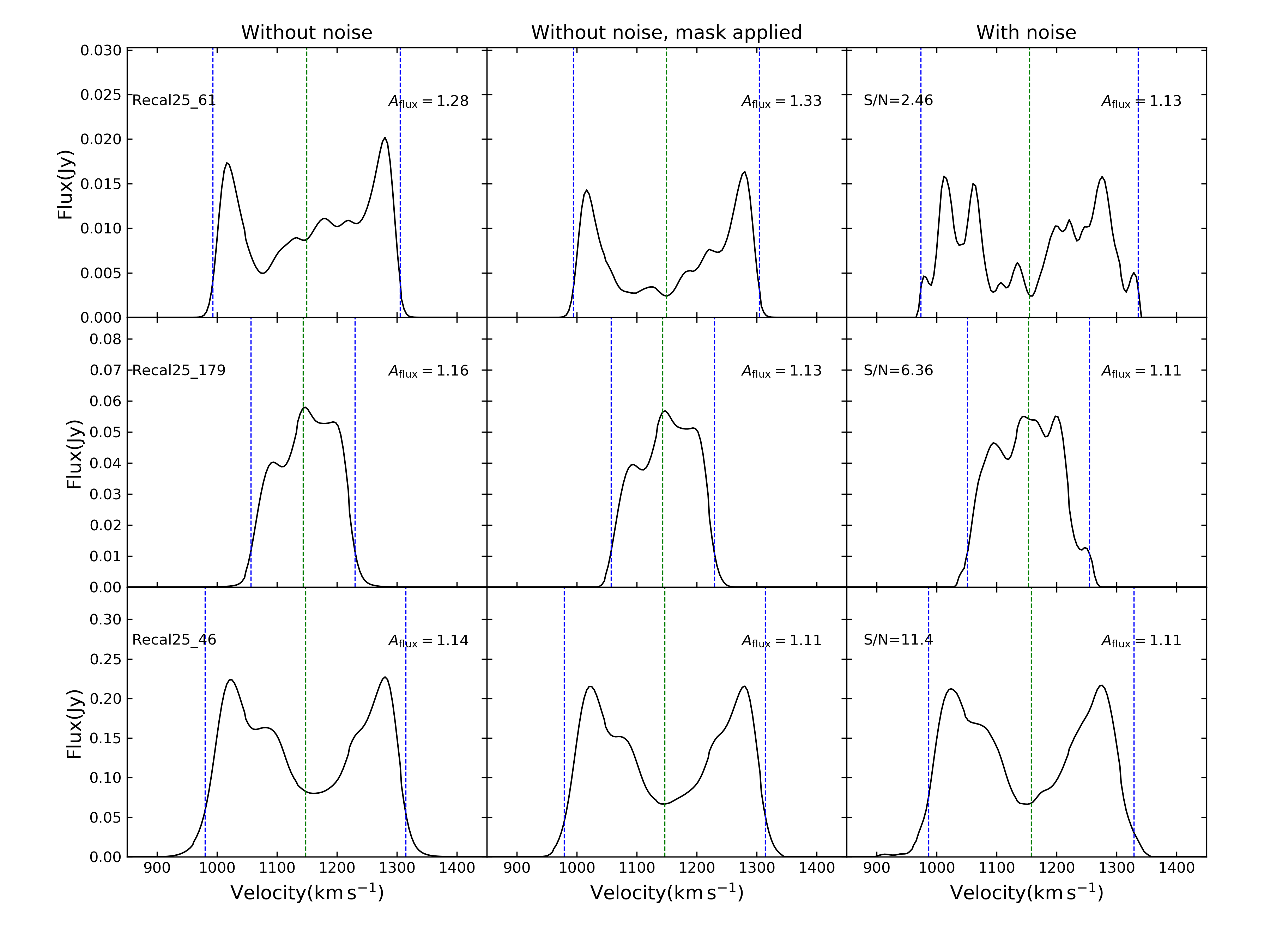}
    \caption{This figure illustrates how noise affects the shape of global profile of mock galaxies at different S/N values. In each panel, the blue dashed lines indicate velocities at 20 per cent of peak flux; the dashed green line indicates the systemic velocity of the global profile. Each row shows noise-free global profile (left panel), mask-applied noise-free global profile (middle panel), and noise-added global profile (right panel) of a mock galaxy. The name of the mock galaxy is shown in the left panel at the top left corner. The S/N value for the global profile after the addition of noise is shown in the right panel at the top left corner. The \Aflux\ value for each version of the global profile is shown at the top right corner of each panel.}
    \label{fig:AfluxMosaic}
\end{figure*}

 The addition of noise varies the shape of the global profile and thereby may increase or decrease its \Aflux\ value. To have a better understanding of the effect of noise, we consider the noise-free global profile, the noise-added global profile, and additionally a mask-applied noise-free global profile where the same 3D mask is applied as in the case with noise at different S/N values. In Fig. \ref{fig:AfluxMosaic}, we illustrate these three versions of global profiles for 3 representative mock galaxies. The noise-added global profile of the mock galaxy `Recal$25\_61$' has a low S/N ($2.46$) and it's shape is significantly different from the shape of the noise-free global profile. It is also evident that several noise peaks appear in different channels of the noise-added global profile, which are absent in the noise-free profile. However, ratio of the noise-added \Aflux\ value to the noise-free \Aflux\ value is still close to unity. In an actual observation, it would be difficult to determine whether the multiple peak structure in the noise-added global profile of `Recal$25\_61$' has an astrophysical origin. The mask-applied noise-free global profile also differs from the noise-free case. This implies that the mask does not enclose all regions of \HI\ emission in the mock datacube. After the addition of noise, a part of the low surface brightness \HI\ gas could lie below the threshold applied during creation of the mask, which then results in an underestimation of flux in the global profile as well as changing its shape.
\par For the other two galaxies, `Recal$25\_179$' and `Recal$25\_46$', the shapes of the noise-free global profile, noise-added global profile and the mask-applied noise-free global profile are similar but not identical. This indicates that the 3D mask encloses most of the regions of \HI\ emission within the mock datacube. Similarly, contribution of noise to the flux in each channel is present but is much less evident for `Recal$25\_46$' due the high S/N of the global profile than for `Recal$25\_179$'. From the behaviour of the \Aflux\ ratio in different S/N bins and the effect of noise on the shape of the noise-added global profile we conclude that a minimum S/N of 5.5 is required for robust measurements of the \Aflux\ value. In global profiles with a lower S/N value, there may be a pronounced change in the shape of the global profile in addition to uncertainties in the \Aflux\ measurement due to noise.
\par 

\section{Correlation between global profile and morphological asymmetry indices}\label{sec:CorrelateAmodAflux}
Ongoing and upcoming \HI\ imaging surveys, such as APERTIF Medium-Deep Survey \footnote{https://www.astron.nl/telescopes/wsrt-apertif/}, MIGHTEE-HI \citep{Mightee2016}, WALLABY \citep{Wallaby2020}, LADUMA \citep{Laduma2016}, and CHILES \citep{Chiles2013}, will have sufficient sensitivity to probe low column density gas down to $5\times10^{19}$\cm\ or less. However, many of the galaxies detected in these surveys beyond the local universe would not be sufficiently resolved to quantify morphological asymmetry and we would have to rely on the global \HI\ profile of the galaxies alone. If a relation between the \Amod\ and \Aflux\ indices exists, a morphological asymmetry may be inferred for spatially unresolved galaxies at higher redshifts.
\par
\begin{figure*}
    \centering
    \includegraphics[width=0.98\textwidth]{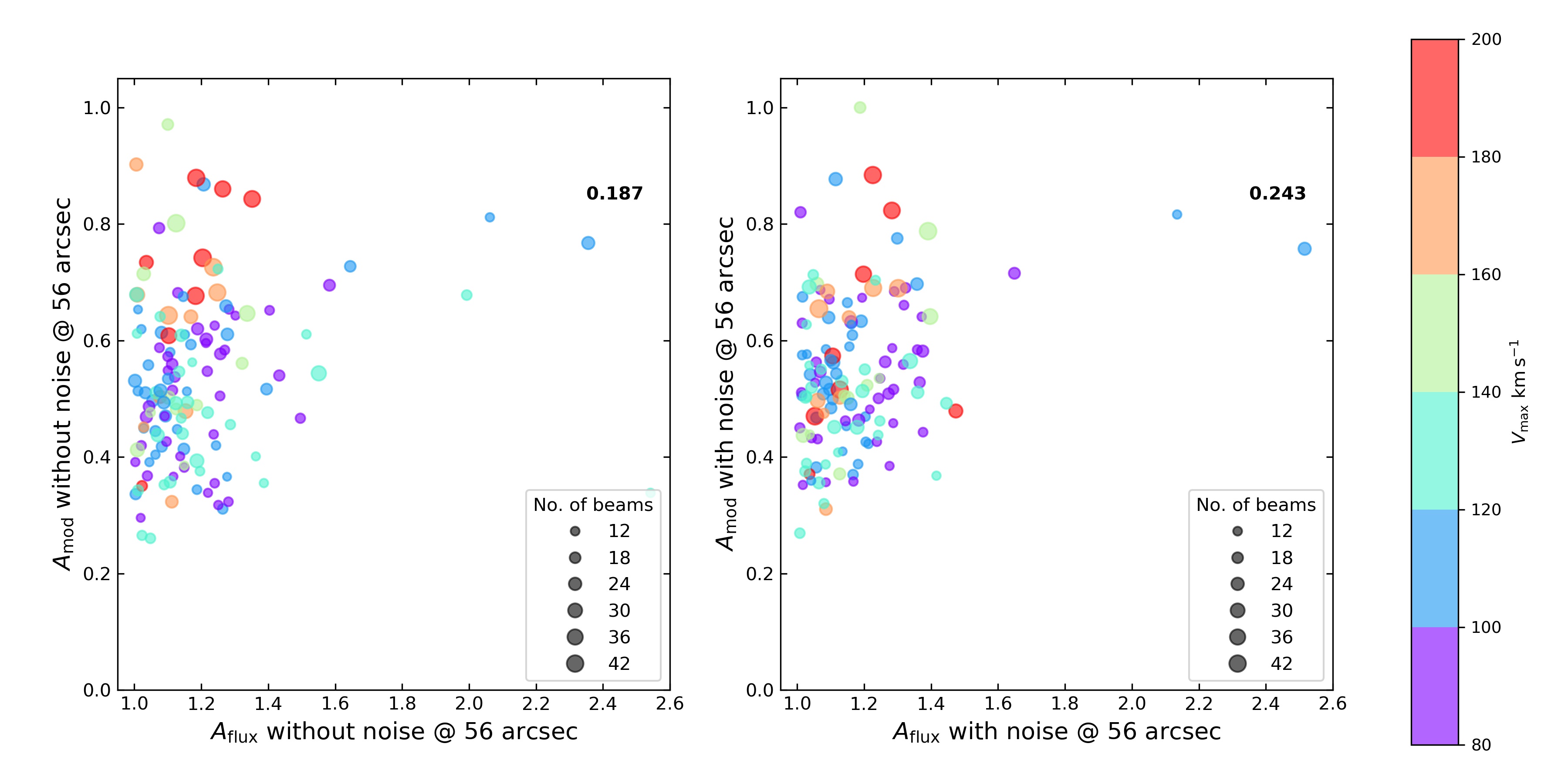}
    \caption{This figure compares asymmetry measured with different indices for the noise-free and noise-added datacubes of mock galaxies. The colors indicate the maximum rotational velocity of galaxies and the size indicates the number of beams across the diagonal of the minimum box of galaxies. The \Aflux\ and \Amod\ values are measured at an angular resolution of 56 arcsec and velocity resolution of $20\,\mathrm{km}\,\mathrm{s}^{-1}$. The left and the right panel show the comparison between the noise-free and noise-added asymmetry values respectively. The \Amod\ values are measured with a column density threshold of $5\times10^{19}$\cm. The Spearman correlation coefficient between the indices is shown at the top-right corner of both panels.}
    \label{fig:CorrelateAmodAflux}
\end{figure*}

 In Fig. \ref{fig:CorrelateAmodAflux}, we illustrate the comparison between the \Amod\ and the \Aflux\ values. In this figure, points are coloured by the $V_\mathrm{max}$ of the galaxies, which closely traces the dynamical mass of the galaxy, while the sizes represent the number of beams across the galaxy. In the left panel we measure the \Amod\ and the \Aflux\ values in noise-free mock datacubes at an angular resolution of 56 arcsec$^{2}$ (4.8 kpc at a distance of 17.1 Mpc) and a velocity resolution of $20\,\mathrm{km}\,\mathrm{s}^{-1}$. In the panel on right, we compare the \Amod\ and the \Aflux\ values in noise-added mock datacubes of the `low noise' sample (discussed in Section \ref{sec:GPasymmetry1}) at the same resolution. Additionally, we only include galaxies that are more than 11 beams resolved and have a global profile S/N$\,>6$. The \Amod\ values were calculated using a column density threshold of $5\times10^{19}$\cm. From Fig. \ref{fig:CorrelateAmodAflux}, we find that galaxies with \Aflux$<1.6$ may have a wide range of \Amod\ values, some even as high as \Amod$\,=0.9$, while all galaxies with \Aflux$>1.6$ tend to have \Amod$\,>0.6$. An inspection of the mock atlas pages and datacubes reveals that these galaxies with high \Aflux\ as well as \Amod\ values seem to be undergoing a merger event where the gas has not yet settled into a regularly rotating disk.
\par In Fig. \ref{fig:CorrelateAmodAflux}, the Spearman correlation coefficient between the noise-free \Amod\ and \Aflux\ is $0.19$, implying that there is no significant relation between the two asymmetry indices. Similarly the correlation coeffcient is $0.24$ in the case of noise-added \Amod\ and \Aflux. Our results are in conflict with the conclusions of \cite{Reynolds2020}, who find a moderate correlation of $\sim0.4$ between the \Aflux\ and the asymmetry index $A$ (see Eq. \ref{Asymm} values of the LVHIS, VIVA, and HALOGAS galaxies). As \cite{Reynolds2020} use the Pearson correlation coefficient and the \Aflux\ values of their sample do not exceed 1.6, we use a subset of our sample with \Aflux$<1.6$ and measure the Pearson correlation coefficient between the \Amod\ and \Aflux\ indices for this subset. We find a correlation coefficient of 0.16 and 0.19 for the noise-free and noise-added asymmetry values, which is still below the values mentioned by \cite{Reynolds2020}. There may be systematic biases (due to the effects described in Section \ref{sec:Moprhasymmetry}) in the $A$ values mentioned by \cite{Reynolds2020} as the column density threshold and the resolution at which these asymmetry values are measured are not mentioned. It is therefore hard to explain why they find a moderate correlation.\\ 
\par As mentioned before in Section \ref{sec:GPAsymmetry}, one reason for the lack of correlation observed in Fig. \ref{fig:CorrelateAmodAflux} could be that the orientation of the galaxy with respect to the observer may result in lower \Aflux\ values even when the underlying \HI\ distribution is asymmetric. \cite{DegHank2020} find that when the inclination is greater than 20\degree, \Aflux\ values are relatively stable against a change in inclination. However, \Aflux\ values have a strong dependence on the position angle of the asymmetric feature in the galaxy. This is because both the \HI\ distribution and the kinematics of a galaxy determine the shape of the global profile. Therefore, asymmetries in the kinematics of a galaxy could also affect the \Aflux\ value. For example, \cite{Swaters1999} study two kinematically lopsided galaxies, DDO 9 and NGC 4395, with fairly symmetric \HI\ distributions. For both DDO 9 and NGC 4395, the rotation curve on the approaching side rises and then flattens while the rotation curve on the receding side continues to rise (see figures 1 and  2 in \cite{Swaters1999}). This introduces an asymmetry in the global profile despite the symmetric \HI\ distribution. The findings of \cite{DegHank2020} and \cite{Swaters1999} suggest that even if a 2D kinematic asymmetry index is considered, we may also not find a relation between the \Aflux\ and the kinematic asymmetry index. Perhaps a 3-D asymmetry index that captures both the kinematic and the morphological asymmetries may have stronger correlation with the \Aflux\ index.
\par
\begin{figure*}
    \centering
    \includegraphics[width=0.8\textwidth]{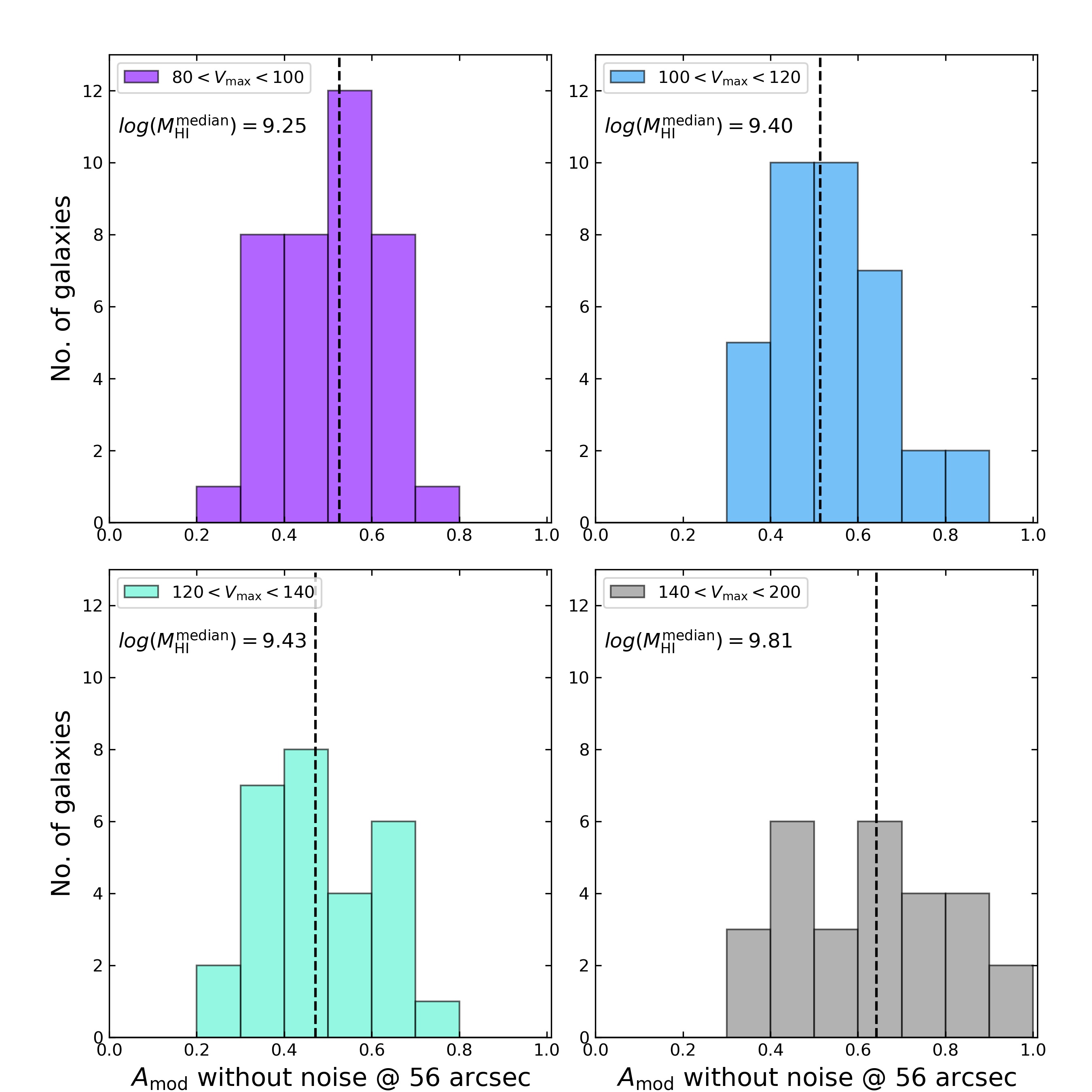}
    \caption{This figure shows the distribution of noise-free \Amod\ values of the mock galaxies in different bins of $V_\mathrm{max}$. The first three bins in $V_\mathrm{max}$ are the same as those shown in Fig. \ref{fig:CorrelateAmodAflux}, while the last three bins are combined into one as each bin has low number of galaxies. The median \HI\ mass of the mock galaxies is shown in the upper left corner of each panel. The dashed black line in each panel indicates the median of the \Amod\ distribution. The \Amod\ values shown here are measured at 56 arcsec resolution and at a column density threshold of $5\times10^{19}$\cm.}
    \label{fig:AmodVmaxHist}
\end{figure*}

 We also examined the dependence of \Aflux\ and \Amod\ values on the $V_\mathrm{max}$ and number of beams across the major axis. We expected galaxies with higher $V_\mathrm{max}$, which are also large galaxies, to have lower \Amod\ values as these galaxies have a deeper potential and thus the gas would be more difficult to disturb. In Fig. \ref{fig:AmodVmaxHist} we show the histogram of \Amod\ values measured in different bins of $V_\mathrm{max}$. The first three panels of Fig. \ref{fig:AmodVmaxHist}, which corresponds the to lowest three $V_\mathrm{max}$ bins in Fig. \ref{fig:CorrelateAmodAflux}, have similar median values. However, galaxies with $V_\mathrm{max}>140\,\mathrm{km}\,\mathrm{s}^{-1}$ have a higher median \Amod\ value, while the overall distribution has also shifted to higher \Amod\ values. \cite{Bahe2016} show that \textsc{eagle }galaxies with log($M_\mathrm{HI}/M_{\odot})>9.5$ have large \HI\ holes in them that result from the feedback recipes in \textsc{eagle}. Galaxies with $V_\mathrm{max}>140\,\mathrm{km}\,\mathrm{s}^{-1}$ have a higher median \HI\ mass which may result in them having large \HI\ holes and consequently higher asymmetry values. \\
\section{Summary and outlook}\label{sec:Conclusion}
In this work, we have studied the effect of observational limitations on global profile and morphological asymmetry indices. For this, we used $189$ mock \HI\ datacubes of galaxies from the \textsc{eagle} hydrodynamical simulations at different angular resolutions and with varied noise levels. \\
\par To quantify the morphological asymmetry of mock galaxies, we used the modified asymmetry parameter \Amod. We analysed column density maps with and without noise, at different resolutions, and with different column density thresholds in which we measured the asymmetries in the outer \HI\ disc of mock galaxies. To detect signatures of environmental processes acting on galaxies, adequate sensitivity to low column densities is required. We found that the correlation between \Amod\ values at a reference column density threshold of $1\times10^{19}$\cm\ to those at higher column density threshold reduced as the applied column density threshold was increased (see Figs. \ref{fig:CompareScatter} and \ref{CompareCorrelation}). Above a threshold of $5\times10^{19}$\cm\ the value of the correlation coefficient in Fig. \ref{CompareCorrelation} falls below 0.5. Therefore, a column density threshold up to $5\times10^{19}$\cm\ is optimal to measure asymmetries in the outer parts of \HI\ disc that are susceptible to environmental processes. At thresholds greater than $5\times10^{19}$\cm\, asymmetries may result from secular evolution processes and thus may be completely unrelated to environmental influence.\\
\par \cite{Giese2016} and \cite{Lelli2014} have shown that lowering the angular resolution of a galaxy lowers its measured asymmetry value. Considering the number of beams across the minimum box of a galaxy as a metric we found that the variation in \Amod\ values is negligible when a galaxy is resolved by more than 25 beams (see Fig. \ref{fig:CompareResolution}). With this as the reference, we measured the relative change in the \Amod\ value at fewer than 25 beams across (see Fig. \ref{fig:ChangeAmodResolution}). At 11 beams across the galaxy, only 10 per cent of the mock galaxies show more than 30 per cent relative change in their \Amod\ values. Therefore, to measure \Amod\ values that are only marginally affected by resolution, at least 11 beams across the galaxy are required.\\
\par The measured morphological asymmetry value of a galaxy generally increases with the addition of noise. To measure \Amod\ values that are relatively unaffected by noise, it is necessary to include pixels in column density maps that have a sufficiently high S/N ratio. We considered column density maps with varying amounts of noise such that the previously motivated threshold of $5\times10^{19}$\cm\ has different S/N values and measured \Amod\ values with this threshold. Comparing the noise-free \Amod\ values of mock galaxies to the noise-added \Amod\ values we found that at S/N$\,>3$, less than 5 per cent of the mock sample shows more than 30 per cent change in the asymmetry value (see Fig. \ref{fig:CompareSN}). Therefore, for \Amod\ values to be minimally unaffected by noise, the chosen threshold should have a S/N of at least 3.
\par In order to ensure a fair comparison of the \Amod\ values among galaxies observed with different resolutions and sensitivities, it is important to uniformly apply the above mentioned constraints. Observational limitations that affect the modified asymmetry index \Amod\ would similarly affect the asymmetry index $A$ as demonstrated by \cite{Giese2016}. Notably, such precautions concerning the measurement of the $A$ values for the WHISP galaxies were not taken by \cite{Holwerda2011} and were shown by \cite{Giese2016} to be unreliable. Similarly, \cite{Reynolds2020} compared the $A$ values of LVHIS, VIVA and HALOGAS galaxies but did not mention details regarding resolution, noise levels, and applied column density threshold. It is conceivable that the $A$ values mentioned in \cite{Reynolds2020} are subject to biases similar to those in \cite{Holwerda2011}.
\par The morphological asymmetry index \Amod\ can identify asymmetric galaxies. However, to identify the underlying physical processes causing these asymmetric features a combination of other non-parametric indices can be used. \cite{Bignone2017} used Gini and M20 indices to identify galaxies undergoing mergers using optical images of galaxies in the ILLUSTRIS simulation. Similarly, a combination of different non-parametric indices can be applied to the \HI\ disc of galaxies in simulations to identify mergers, tidal interactions, and ram pressure stripping in galaxies. This will be the subject of a forthcoming study.
\par We characterise asymmetries in the global profiles of mock galaxies by measuring the \Aflux\ index. We use the \Aflux\ distribution of the AMIGA sample of isolated galaxies as a reference and we find that 9.8 per cent of the mock galaxies have \Aflux$\,>1.39$ when measured without noise. For the two sets of noise-added global profiles we find that 7 per cent and 32 per cent of mock global profiles have \Aflux$\,>1.39$ in the `high S/N' and `low S/N' subsets respectively. This implies that the mock galaxies from \textsc{eagle} simulations are intrinsically more asymmetric than the AMIGA sample. In global profiles with S/N$\,<5.5$, the addition of noise may result in a large change in \Aflux\ as compared to the noise-free \Aflux\ value (see Fig. \ref{fig:AfluxCompare} and Table \ref{tab:GPAsymValues1}). Similarly, the shape of the global profile may also change substantially, which may or may not affect the \Aflux\ measurements (see Fig. \ref{fig:AfluxMosaic}).
\par We investigate the relation between the \Aflux\ and \Amod\ indices using mock datacubes and find a small correlation coefficient of $\sim0.2$ for both the noise-free and noise-added asymmetry values (see Fig. \ref{fig:CorrelateAmodAflux}). This lack of correlation could result from the effect of the kinematics of a galaxy on the shape of the global profile. The absence of a relation between the two asymmetry parameters implies that a morphological asymmetry cannot be inferred from the global profile asymmetry value of spatially unresolved galaxies.
\par As a next step, the constraints for calculating the morphological asymmetry index derived in this work will be applied to observations of real galaxies from \HI\ imaging surveys of the Ursa Major and the Perseus-Pisces volumes. By quantifying morphological asymmetries in the \HI\ disc of galaxies, we will investigate the effects of environmental processes prevalent in these volumes.
\section*{Data Availability}
The data underlying this article will be shared on reasonable request to the corresponding author.
\section*{Acknowledgements}
We would like to thank the reviewer for their careful consideration of our manuscript. Their suggestions have helped improve the text greatly.
\par PB acknowledges the Leids-Kerhoven Bosscha Fund for financial travel support. KAO acknowledges support by the European Research Council (ERC) through Advanced Investigator grant to C.S.Frenk, DMIDAS (GA 786910) and by the Netherlands Foundation for Scientific Research (NWO) through VICI grant 016.130.338 to MV. PB and MV acknowledges support by the Netherlands Foundation for Scientific Research (NWO) through VICI grant 016.130.338. JMvdH acknowledges support from the European Research Council under the European Union's Seventh Framework Programme (FP/2007-2013) / ERC Grant Agreementnr. 291531 (HIStoryNU). 
\par We acknowledge the Virgo Consortium for making their simulation data available. The \textsc{eagle} simulations were performed using the DiRAC-2 facility at Durham, managed by the ICC, and the PRACE facility Curie based in France at TGCC, CEA, Bruy\`eresle-Ch\^atel. 
\par This work used the DiRAC@Durham facility managed by the Institute for Computational Cosmology on behalf of the STFC DiRAC HPC Facility (www.dirac.ac.uk). The equipment was funded by BEIS capital funding via STFC capital grants ST/K00042X/1, ST/P002293/1, ST/R002371/1 and ST/S002502/1, Durham University and STFC operations grant ST/R000832/1. DiRAC is part of the National e-Infrastructure.



\bibliographystyle{mnras}
\bibliography{Bibentries.bib} 




\appendix
\section{Properties of mock galaxies}\label{appendix:A}
In Table \ref{table:MockProperties} we present the group numbers of mock galaxies from the Recal25 run of the \textsc{eagle} simulations and properties derived from the mock datacubes for each galaxy. We select central galaxies from each FoF group at $z=0$ with maximum rotational velocities ($V_\mathrm{max}$) in the range $80\,\mathrm{km}\,\mathrm{s}^{-1}<V_\mathrm{max}<200\,\mathrm{km}\,\mathrm{s}^{-1}$. Thus, each galaxy mentioned in this table has SubGroupNumber = 0 and SnapNum= 28 in the SubHalo table of Recal25 in the \textsc{eagle} database (see http://icc.dur.ac.uk/Eagle/database.php and \citealt{EagleDataRelease}). The full table is available as supplementary online material.
\begin{table*}
    \centering
    \caption{Properties of the mock sample of galaxies extracted from the \textsc{eagle} simulations. Columns: \textbf{(1)} GroupNumber of the central galaxy; \textbf{(2)} Maximum rotational velocity; \textbf{(3)} \HI\ mass; \textbf{(4)} Inclination; \textbf{(5) to (9)} \Amod\ value measured in noise-free column density maps with a threshold of $5\times10^{19}\,\mathrm{cm}^{-2}$ at resolutions of $12\,\mathrm{arcsec}\times17\,\mathrm{arcsec}$, $30,45,56,\text{ and }98\,\mathrm{arcsec}$ respectively; \textbf{(10)} \Aflux\ value measured in $56\,\mathrm{arcsec}$ noise-free global profiles. Only a few rows are shown here, the full table can be found as supplementary online material.}
    \begin{tabular}{c c c c c c c c c c}
    \hline
    Group Number & $V_\mathrm{max}$ & log($M_\mathrm{HI}$) & Inclination & $A_\mathrm{mod}^\mathrm{nf}$ & $A_\mathrm{mod}^\mathrm{nf}$ & $A_\mathrm{mod}^\mathrm{nf}$ & $A_\mathrm{mod}^\mathrm{nf}$ & $A_\mathrm{mod}^\mathrm{nf}$ &$A_\mathrm{flux}^\mathrm{nf}$\\
      (Recal25\_) & ($\mathrm{km}\,\mathrm{s}^{-1})$ & ($\text{M}_{\odot}$) & (degrees) & (@12 arcsec) & (@30 arcsec) & (@45 arcsec) & (@56 arcsec) & (@98 arcsec) & (@56 arcsec)\\
    \hline
    17 & 205 & 9.98 & 49 & 0.82 & 0.77 & 0.72 & 0.68 & 0.49 & 1.18\\ 
	25 & 199 & 9.98 & 65 & 0.90 & 0.89 & 0.86 & 0.84 & 0.77 & 1.35\\ 
	27 & 185 & 9.53 & 66 & 0.95 & 0.93 & 0.90 & 0.86 & 0.50 & 1.26\\ 
	28 & 198 & 9.86 & 93 & 0.84 & 0.81 & 0.77 & 0.74 & 0.54 & 1.20\\ 
	30 & 193 & 9.22 & 108 & 0.88 & 0.83 & 0.78 & 0.73 & 0.60 & 1.04\\ 
	31 & 189 & 10.00 & 65 & 0.73 & 0.69 & 0.65 & 0.61 & 0.53 & 1.10\\ 
	32 & 183 & 9.96 & 41 & 0.90 & 0.89 & 0.89 & 0.88 & 0.85 & 1.19\\ 
	33 & 166 & 10.25 & 83 & 0.77 & 0.75 & 0.74 & 0.73 & 0.69 & 1.24\\ 
	34 & 174 & 10.10 & 111 & 0.74 & 0.71 & 0.67 & 0.64 & 0.55 & 1.10\\ 
	35 & 178 & 9.76 & 19 & 0.65 & 0.59 & 0.52 & 0.48 & 0.37 & 1.15\\ 
	36 & 180 & 9.42 & 75 & 0.48 & 0.42 & 0.37 & 0.35 & 0.29 & 1.02\\ 
	37 & 158 & 9.87 & 17 & 0.85 & 0.83 & 0.81 & 0.80 & 0.74 & 1.12\\ 
	38 & 167 & 9.81 & 115 & 0.63 & 0.58 & 0.53 & 0.50 & 0.43 & 1.08\\ 
	39 & 141 & 9.85 & 73 & 0.77 & 0.75 & 0.73 & 0.71 & 0.67 & 1.03\\ 
	40 & 166 & 9.95 & 44 & 0.73 & 0.70 & 0.67 & 0.64 & 0.56 & 1.17\\  
    \hline
    \end{tabular}
    \label{table:MockProperties}
\end{table*}

\section{Atlas pages of mock galaxies}\label{appendix:B}
\begin{figure*}
    \centering
    \includegraphics[angle=90, trim={1.8cm 0cm 0cm 0cm }, clip, width=0.7\textwidth]{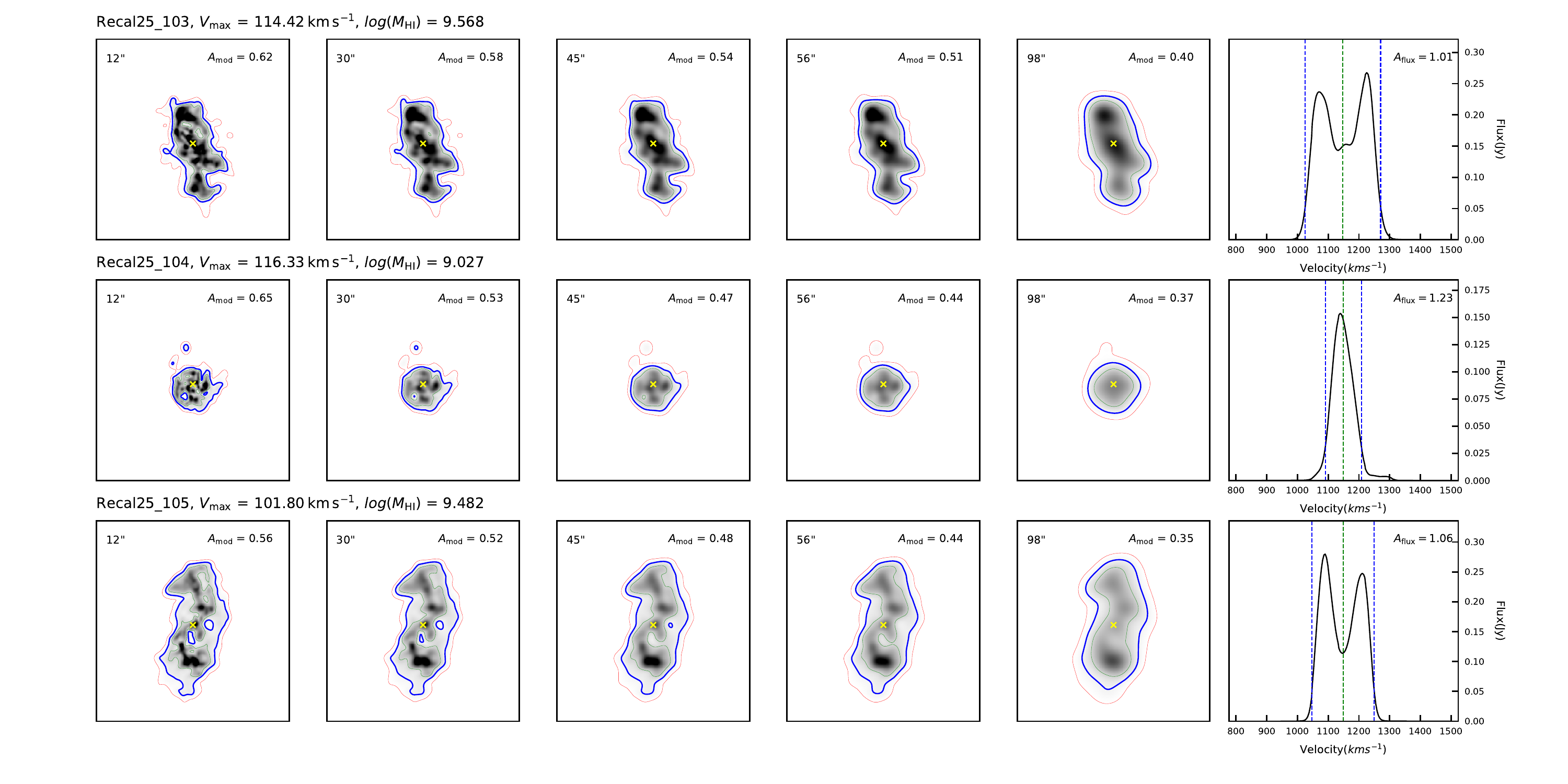}
    \caption{An example of atlas page, see text of Appendix B for detail}
    \label{fig:AtlasExample}
\end{figure*}

Fig \ref{fig:AtlasExample} shows an example atlas page, where each page has data products of three mock galaxies. For each galaxy, we present noise-free \HI\ data products derived in Section \ref{MockHICubes}. The FoF group number of the galaxy, it's maximum rotational velocity and logarithm of \HI\ mass is shown above each row. We present column density maps at angular resolutions of $12\times17$, 30, 45, 56 and 98 arcsec in which column densities of 1, 5, and 15 $\times10^{19} $\cm\ are shown as red, blue and green contours respectively. Each panel showing column density maps is 100 kpc across. At each angular resolution, the \Amod\ value measured at a column density threshold of $5\times10^{19}$\cm\ is shown in the top right corner. The center around which the galaxy is rotated is shown with a yellow cross. A noise-free global profile of the galaxy derived at 56 arcsec is shown, where blue dashed lines indicate velocities at 20 per cent of the peak flux and the green dashed line indicates the systemic velocity. The \Aflux\ of the galaxy is shown in the top right corner. The atlas pages are available in full as supplementary online material.

\bsp	
\label{lastpage}
\end{document}